\documentclass[journal]{journal}
\pdfoutput=1
\usepackage{multicol,lipsum}
\usepackage{booktabs}
\usepackage[normalem]{ulem}
\usepackage{longtable}
\useunder{\uline}{\ul}{}

\ifCLASSINFOpdf
	\usepackage[pdftex]{graphicx}
 
\else
\fi
\usepackage[cmex10]{amsmath}

\usepackage{mathtools}

\usepackage{float}

\usepackage[export]{adjustbox}

\usepackage{fixltx2e}
\usepackage{ulem}
\usepackage{subfigure}
\usepackage{xcolor}
\usepackage{setspace}

\usepackage[colorlinks,
            linkcolor=blue,
            anchorcolor=blue,
            citecolor=blue]{hyperref}

\usepackage{nameref}

\begin{document}
%
\title{Newtonian Mechanics Based Transient Stability PART III: Superimposed Machine}
%
%
%

\author{{Songyan Wang,
        Jilai Yu,  
				Aoife Foley,
        Jingrui Zhang
        }
        
}
%
%

\markboth{Journal of \LaTeX\ Class Files,~Vol.~6, No.~1, January~2007}%
{Shell \MakeLowercase{\textit{et al.}}: Bare Demo of IEEEtran.cls for Journals}
%



\maketitle
\thispagestyle{empty}
\begin{abstract}

This paper analyzes the mechanisms of the superimposed machine and also its inherit problems in TSA. Based on the global monitoring of the original system trajectory, the transient energy is mistakenly defined as the superimposition of the transient energy of all machines in the system.
This ``energy superimposition” directly causes the superimposed machine to become a pseudo machine without any equation of motion, and in this way the superimposed machine completely violates all the machine paradigms. The violations bring the two inherit defects in TSA: (i) the stability of the superimposed machine is unable to be characterized precisely, and (ii) the variance of the original system trajectory is unstable to be depicted clearly. The two defects are also reflected in the definitions of the superimposed-machine based transient stability concepts.
In particular, the swing and the critical stability of the system are unable to be defined strictly, and the potential energy surface cannot be modeled precisely. Simulation results show that the problems of the pseudo superimposed-machine in TSA.

\end{abstract}

%

\IEEEpeerreviewmaketitle
\begin{small}

\begin{tabular}{lllll}
    &            &               &                  &                                \\
  \multicolumn{5}{c}{\textbf{Nomenclature}}                                                \\

  KE     & \multicolumn{1}{c}{}  & \multicolumn{3}{l}{Kinetic energy}                      \\
  PE     & \multicolumn{1}{c}{}  & \multicolumn{3}{l}{Potential energy}                    \\
  COI    & \multicolumn{1}{c}{}  & \multicolumn{3}{l}{Center of inertia}                   \\
  DLP    &                       & \multicolumn{3}{l}{Mechanics liberation point}            \\
  DSP    &                       & \multicolumn{3}{l}{Mechanics stationary point}            \\
  EAC    &                       & \multicolumn{3}{l}{Equal area criterion}                \\
  MPP    &                       & \multicolumn{3}{l}{Maximum potential energy point}      \\
  NEC    &                       & \multicolumn{3}{l}{Newtonian energy conversion}         \\
  TSA    &                       & \multicolumn{3}{l}{Transient stability assessment}      \\
  TEF    &                       & \multicolumn{3}{l}{Transient energy function}           \\
  GTE    &                       & \multicolumn{3}{l}{Global total transient energy}           \\
  GKE    &                       & \multicolumn{3}{l}{Global KE}           \\
  GPE    &                       & \multicolumn{3}{l}{Global PE}           \\ 
  PEF    &                       & \multicolumn{3}{l}{Partial energy function}           \\
  PEB    &                       & \multicolumn{3}{l}{Potential energy boundary}           \\
  PES    &                       & \multicolumn{3}{l}{Potential energy surface}           \\
  UEP    &                       & \multicolumn{3}{l}{Unstable equilibrium point}           \\
  EEAC    &                       & \multicolumn{3}{l}{Extended EAC}           \\
  GPEB    &                       & \multicolumn{3}{l}{Global PEB}           \\
  GPES    &                       & \multicolumn{3}{l}{Global PES}           \\
  GMPP    &                       & \multicolumn{3}{l}{Global MPP}           \\
  LOSP    &                       & \multicolumn{3}{l}{Loss of synchronism point}           \\ 
  IMEF    &                       & \multicolumn{3}{l}{Individual-machine energy function}           \\ 
  IMKE    &                       & \multicolumn{3}{l}{Individual-machine KE}           \\ 
  IMPE    &                       & \multicolumn{3}{l}{Individual-machine PE}           \\ 
  IMPP    &                       & \multicolumn{3}{l}{Individual-machine MPP}           \\ 
  IMTE    &                       & \multicolumn{3}{l}{Individual-machine total transient energy}           \\ 
  IMTR    &                       & \multicolumn{3}{l}{Individual-machine trajectory}           \\ 
  IVCS    &                       & \multicolumn{3}{l}{Individual-machine-virtual-COI-machine system}           \\ 
  RUEP    &                       & \multicolumn{3}{l}{Relevant UEP}           \\  
  IEEAC    &                       & \multicolumn{3}{l}{Integrated EEAC}           \\  
  IMEAC    &                       & \multicolumn{3}{l}{Individual-machine EAC}           \\  
  IMPEB    &                       & \multicolumn{3}{l}{Individual-machine PEB}           \\  
  IMPES   &                       & \multicolumn{3}{l}{Individual-machine PES}   \\

\end{tabular}
\end{small}

\section{Introduction} \label{section_I}

%
%
%
%
\subsection{LITERATURE REVIEW} \label{section_IA}
\raggedbottom
\IEEEPARstart{S}{trict} followings of the machine paradigms ensure the advantages of the individual-machine in TSA \cite{1}, \cite{2}.
However, if we retrospect the previous transient stability studies in history, an intuitive idea is that the transient energy of all machines in the system should be ``superimposed”. 
Against this background, the ``one-and-only” global machine with the superimposed transient energy is established. The Transient energy conversion inside this one-and-only superimposed machine is also believed to be the representation of the entire system. Both the RUEP method \cite{3} and the sustained fault method \cite{4}, \cite{5} were fully based on this superimposed-machine thinking.
\par Superimposed-machine is of value in the history of the transient stability assessment because it released the restriction of the Lyapunov theories. In particular, the famous Newtonian scenario that an energy ball rolling in basin was first established through the superimposed-machine potential energy surface \cite{3}-\cite{5}.
Essentially speaking, the key concepts and fundamental theories in the modern transient stability analysis are inspired by the Newtonian-mechanical thinking in the early superimposed-machine studies. Unfortunately, because of the misunderstanding of the original system trajectory, the SMTE is mistakenly defined in a superimposed manner. At the first glance, it seems that the transient characteristic of each individual-machine is still preserved in TSA.
However, the system engineer actually observes the conversion between superimposed-machine kinetic energy (SMKE) and superimposed-machine potential energy (SMPE) rather than that inside each physically real machine. Under this background, a most troublesome problem occurs. That is, the NEC of the superimposed machine completely fails because the residual SMKE always exists no matter the system maintains stable or not \cite{6}, \cite{7}.
The failure of NEC directly causes the confusions of the definitions of the SMTE-based transient stability concepts. The inherit problems of the superimposed machine were never completely solved until nowadays.
\par Based on the machine paradigms as analyzed in the previous papers \cite{1}, \cite{2}, and also the definition of the superimposed machine, two questions can be emerged as follows:
\\ (i) Could the transient characteristics of the superimposed-machine be explained from the individual-machine perspective?
\\ (ii) Could inherit problems in the SMTE be explained from the perspective of machine paradigms?
\par Obviously, answering the two questions may help readers take a deep insight into the mechanisms of the superimposed machine from the angle of ``energy superimposition”.

\subsection{SCOPE AND CONTRIBUTION OF THE PAPER} \label{section_IB}

This paper focuses on the explanations of the mechanisms of the superimposed-machine and it's inherit problems in TSA through the ``energy superimposition” of the individual-machine.
\par Because of the global monitoring of the original system trajectory, the transient energy is mistakenly defined in a superimposed manner. Then, the superimposed machine becomes a pseudo machine without equation of motion. The transient characteristics of the superimposed machine are explained from the individual-machine perspective. It is clarified that the pseudo superimposed-machine complete violates all the machine paradigms. These violations bring the two inherit defects in TSA:
(i) the stability of the superimposed machine is unable to be characterized precisely, and (ii) the variance of the original system trajectory is unstable to be depicted clearly. Caused by the two defects, the trajectory variance and the superimposed-machine transient energy conversion does not match. After that, the two defects of the superimposition are also reflected in the definitions of the transient stability concepts. In particular, the swing of the original system is unable to be depicted clearly (trajectory-depiction defect), the critical stability of the original system cannot be characterized precisely (the two defects), while the superimposed-machine potential energy surface cannot be modeled precisely (the stability-characterization advantage). Simulation results show the problems of the superimposed-machine in TSA.
\par The contributions of this paper are summarized as follows:
(i) The transient characteristics of the superimposed machine are systematically explained through the individual-machine perspective. This explains the mechanism of the superimposed machine from an individual machine manner.
\\ (ii) The superimposed machine transient stability is analyzed. This exposes the defects of the superimposed machine.
\\ (iii) All the transient stability concepts show defects once they are defined based on the superimposed machine. This further validates the reasonability of the machine paradigms.
\par The reminder of the paper is organized as follows. In Section \ref{section_II}, the mechanisms of the superimposed-machine are analyzed. In Section \ref{section_III}, the characteristics of the superimposed machine are explained in an individual-machine manner.
In Section \ref{section_IV}, the superimposed-machine based original system stability is given through the violations of the machine paradigms. In Section \ref{section_V}, the defects of the superimposed-machine based transient stability concepts are analyzed through the violations of the machine paradigms. 
In Section \ref{section_VI}, simulation cases show the problems of the superimposed-machine in TSA. In Section \ref{section_VII}, the third defect of UEP is exposed through its approximation of the superimposed machine based critical transient energy. Conclusions are given in Section \ref{section_VIII}.\\\\

\section{MECHANISMS OF THE SUPERIMPOSED-MACHINE} \label{section_II}
\subsection{GLOBAL MONITORING}  \label{section_IIA}

In the original system trajectory \cite{1}, \cite{2}, the IMTR of each machine is denoted as
\begin{equation}
  \label{equ1}
  \delta_{i\mbox{-}\mathrm{SYS}}=\delta_i-\delta_{\mathrm{SYS}}
\end{equation}
\par However, different from the individual-machine perspective that monitors the IMTR of each machine in the original system trajectory as in Ref. \cite{2}, the superimposed-machine analyst monitors the original system trajectory in a ``global” manner. That is, the entire original system trajectory is seen as a ``whole”.
\par Comparisons between individual-machine monitoring and the global monitoring are shown in Figs. \ref{fig1} (a) and (b), respectively.

\begin{figure} [H]
  \centering 
  \subfigure[]{%
  \label{fig1a}
    \includegraphics[width=0.42\textwidth]{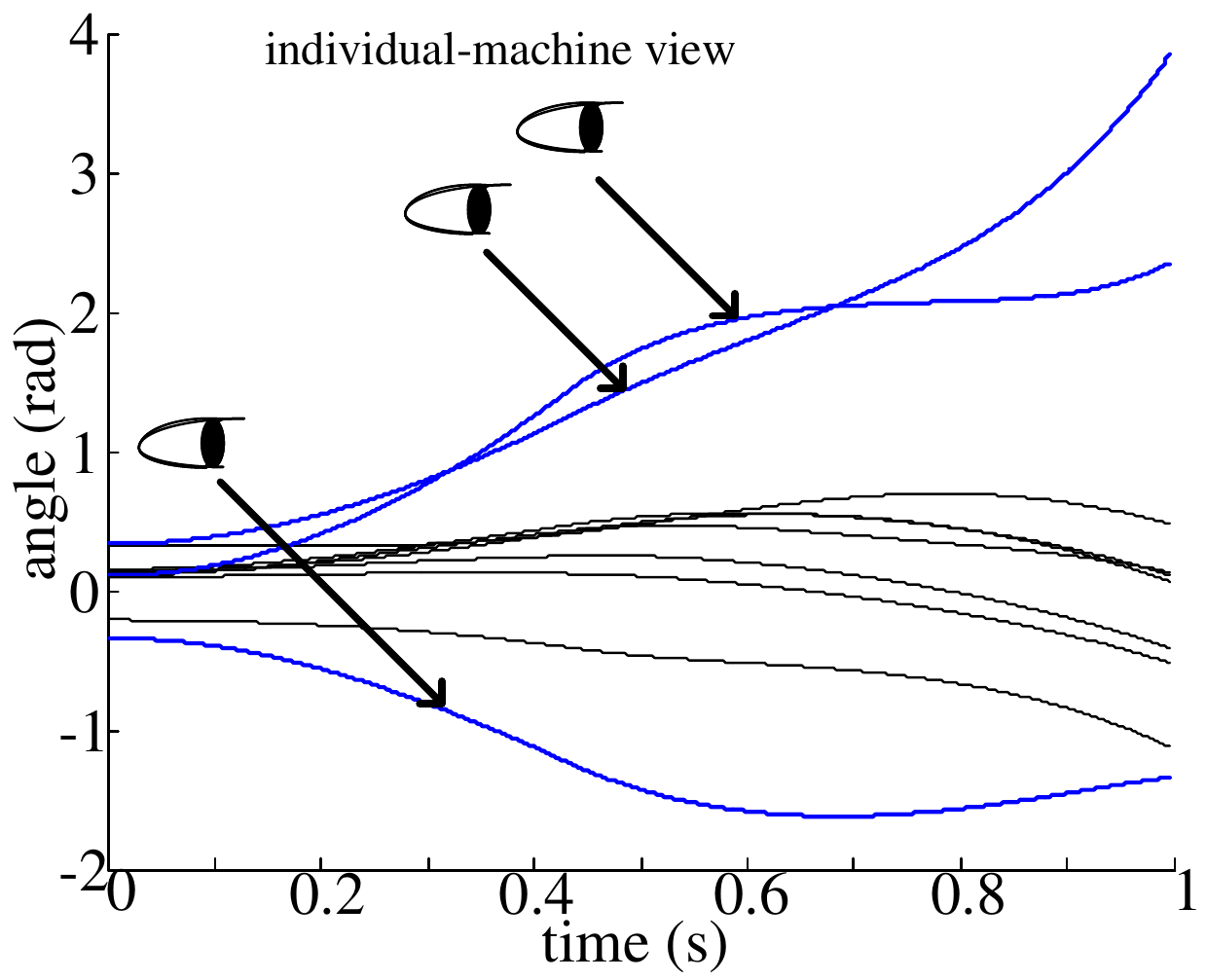}}%
\end{figure} 
\addtocounter{figure}{-1}
\vspace*{-2em}       
\begin{figure} [H]
  \addtocounter{figure}{1}      
  \centering 
  \subfigure[]{%
    \label{fig1b}
    \includegraphics[width=0.42\textwidth]{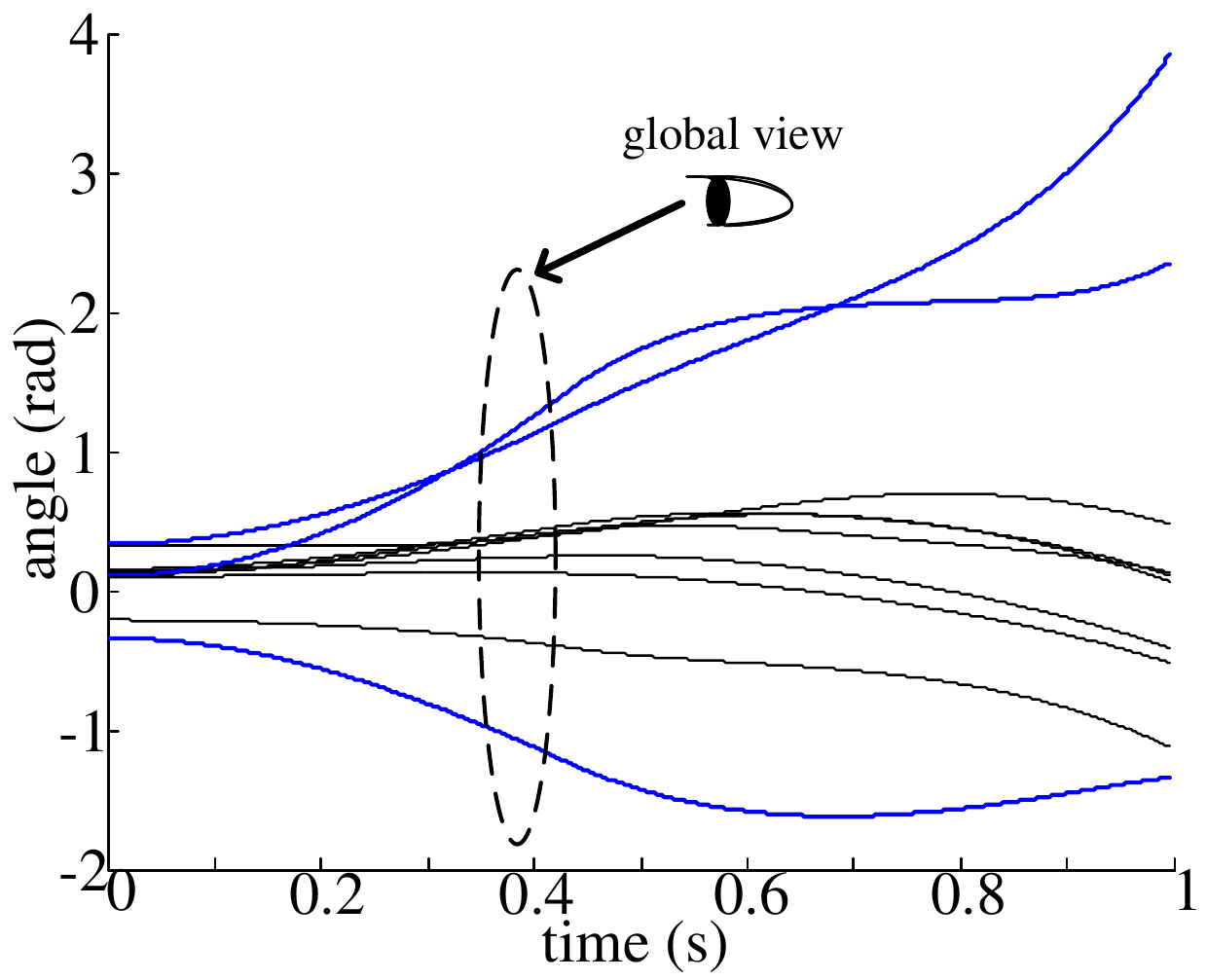}}%
  \caption{Comparisons between individual-machine monitoring and global monitoring [TS-1, bus-2, 0430 s]. (a) Individual-machine angle. (b) Global angle.}%
  \label{fig1}
\end{figure}
\vspace*{-0.5em}
From Fig. \ref{fig1b}, the global monitoring still focuses on the original system trajectory, and it is also based on the COI-SYS reference.
However, because the IMTRs of all machines are treated as a ``whole”, the ``separation” of an objective machine with respect to the COI-SYS cannot be found under this circumstance. Against this background, the key effect of the critical machine to the stability of the original system is completely neglected.

\subsection{FAILURE OF TWO-MACHINE MODELING} \label{section_IIB}
Because of the global monitoring, the ``separation” of an objective machine with respect to the COI-SYS is unable to be established.  Then, the two-machine system is unable to be modeled. Under this background, we still assume a machine exists in the system. However, this machine is ``pseudo” because it does not have equation of motion. That is, this machine does not have trajectory, mass and force on it. This ``pseudo” machine only has energy conversion.

\subsection{SUPERIMPOSED MACHINE TRANSIENT ENERGY CONVERSION} \label{section_IIC}
Based on the global monitoring of the original system trajectory, the transient energy is also defined in a superimposition manner. The superimposed-machine transient energy (SMTE) is denoted as \cite{6}, \cite{7}
\begin{equation}
  \label{equ2}
  V_ {\mathrm{SM} \mbox{-} \mathrm{SYS} }=V_{K E \mathrm{SM}\mbox{-}\mathrm{SYS}}+V_{P E \mathrm{SM}\mbox{-}\mathrm{SYS}}=\sum_{i=1}^{n} V_{i\mbox{-}\mathrm{SYS}}
\end{equation}
where \vspace{0.5em}

\noindent
$\left\{\begin{array}{l}
V_{KE\mathrm{SM}\mbox{-}\mathrm{SYS}}=\sum_{i=1}^{n} V_{K E i\mbox{-}\mathrm{SYS}} \\
\\
V_{P E \mathrm{SM}\mbox{-}\mathrm{SYS}}=\sum_{i=1}^{n} V_{P E i\mbox{-}\mathrm{SYS}}
\end{array}\right.$ \vspace{0.5em}
\par In Eq. (\ref{equ2}), SMKE and SMPE are defined as the superimposition of IMKEs and IMPEs of all machines in the system, respectively.
\par The residual SMKE at the SMPP is denoted as
\begin{equation}
  \begin{split}
    V_{K E \mathrm{SM}\mbox{-}\mathrm{S Y S}}^{R E}&=\sum_{i=1}^{n} V_{K E i\mbox{-}\mathrm{SYS}}^{\mathrm{SMPP}} \\
    &=V_{K E \mathrm{SM}\mbox{-}\mathrm{SYS}}^{c}-\Delta V_{P E \mathrm{SM}\mbox{-}\mathrm{SYS}}
  \end{split}
  \label{equ3}
\end{equation}
where
\begin{spacing}{1.5}
  \noindent $V_{K E \mathrm{SM}\mbox{-}\mathrm{SYS}}^{c}=\sum_{i=1}^{n} V_{K E i\mbox{-}\mathrm{SYS}}^{c}$ \\
  $\Delta V_{P E \mathrm{SM}\mbox{-}\mathrm{SYS}}=\sum_{i=1}^{n} V_{P E i\mbox{-}\mathrm{SYS}}^{\mathrm{SMPP}}-\sum_{i=1}^{n} V_{P E i\mbox{-}\mathrm{SYS}}^{c}$
\end{spacing}
In Eq. (\ref{equ3}), the SMKE and SMPE are depicted as the superimpositions of the IMKEs and IMPEs of all machines in the system, respectively.
\par In this paper, we name the pseudo machine with SMTE the ``superimposed machine”. The superimposed machine only has transient energy conversion. In addition, this energy conversion does not satisfy NEC characteristic. That is, the concepts of DSP and DLP do not exist in the pseudo superimposed machine. Detailed analysis about the pseudo superimposed machine will be given in Section \ref{section_IV}.
\par
\begin{figure}[H]
  \centering
  \includegraphics[width=0.4\textwidth,center]{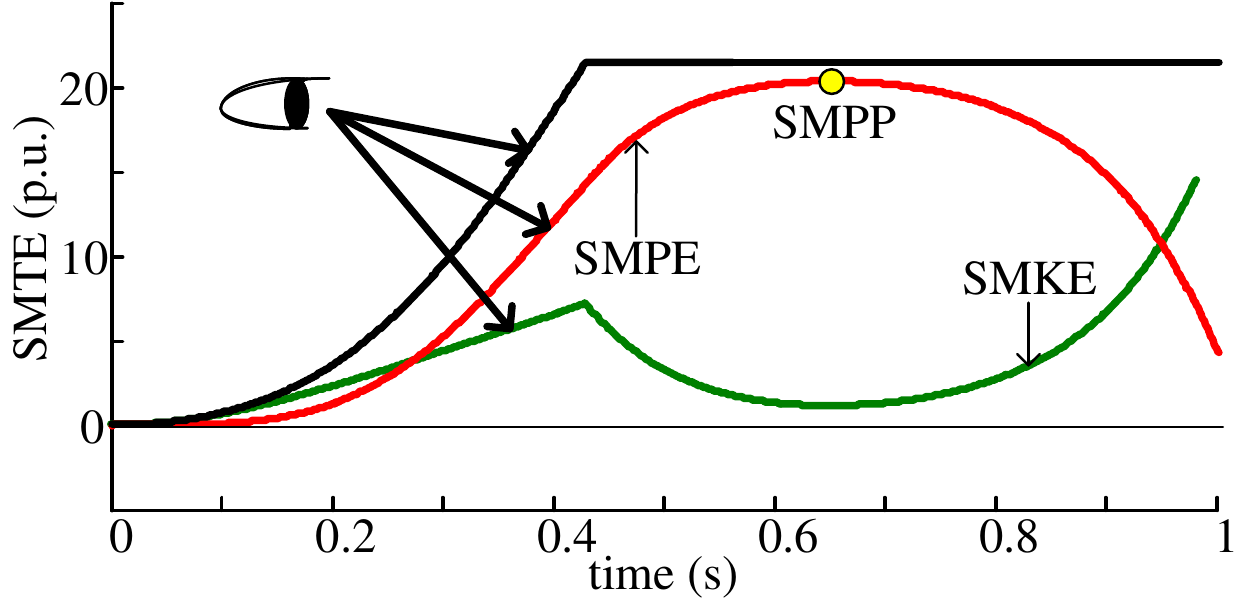}
  \caption{Superimposed machine transient energy conversion [TS-1, bus-2, 0430 s].} 
  \label{fig2}  
\end{figure}

 The superimposed-machine transient energy conversion is shown in Fig. \ref{fig2}.
\vspace*{-0.5em}
From Fig. \ref{fig2}, the superimposed-machine is the ``one-and-only” machine in TSA because SMTE is defined based on the superimposition of all IMTEs in the system.
The superimposed-machine analysts only monitor the conversion between SMKE and SMPE. In fact, although the IMTE ``mathematically” exists in the SMTE, it ``actually” is not used in TSA.
\par The characteristics of the SMTE are given as below.
\vspace*{0.5em}
\\ (Characteristic-I) SMTE keeps conserved once fault is cleared.
\\ (Characteristic-II) SMPP occurs during the post-fault period.
\\ (Characteristic-III) SMPE goes negative infinite along time horizon if the system becomes unstable.
\\ (Characteristic-IV) Residual SMKE always exists no matter the system maintains stable or not.
\vspace*{0.5em}
\par Because the SMTE is defined based on the superimposition of all the IMTEs in the system, in the following section, all the characteristics above will be explained in an individual-machine manner. This may help readers deeply understand the ``energy superimposition” of the superimposed machine.

\section{EXPLANATION OF THE SUPERIMPOSED MACHINE FROM INDIVIDUAL-MACHINE PERSPECTIVE} \label{section_III}
\subsection{CONSERVATIVENESS OF SMTE (CHARACTERISTIC-I)} \label{section_IIIA}
\begin{figure}[H]
  \centering
  \includegraphics[width=0.41\textwidth,center]{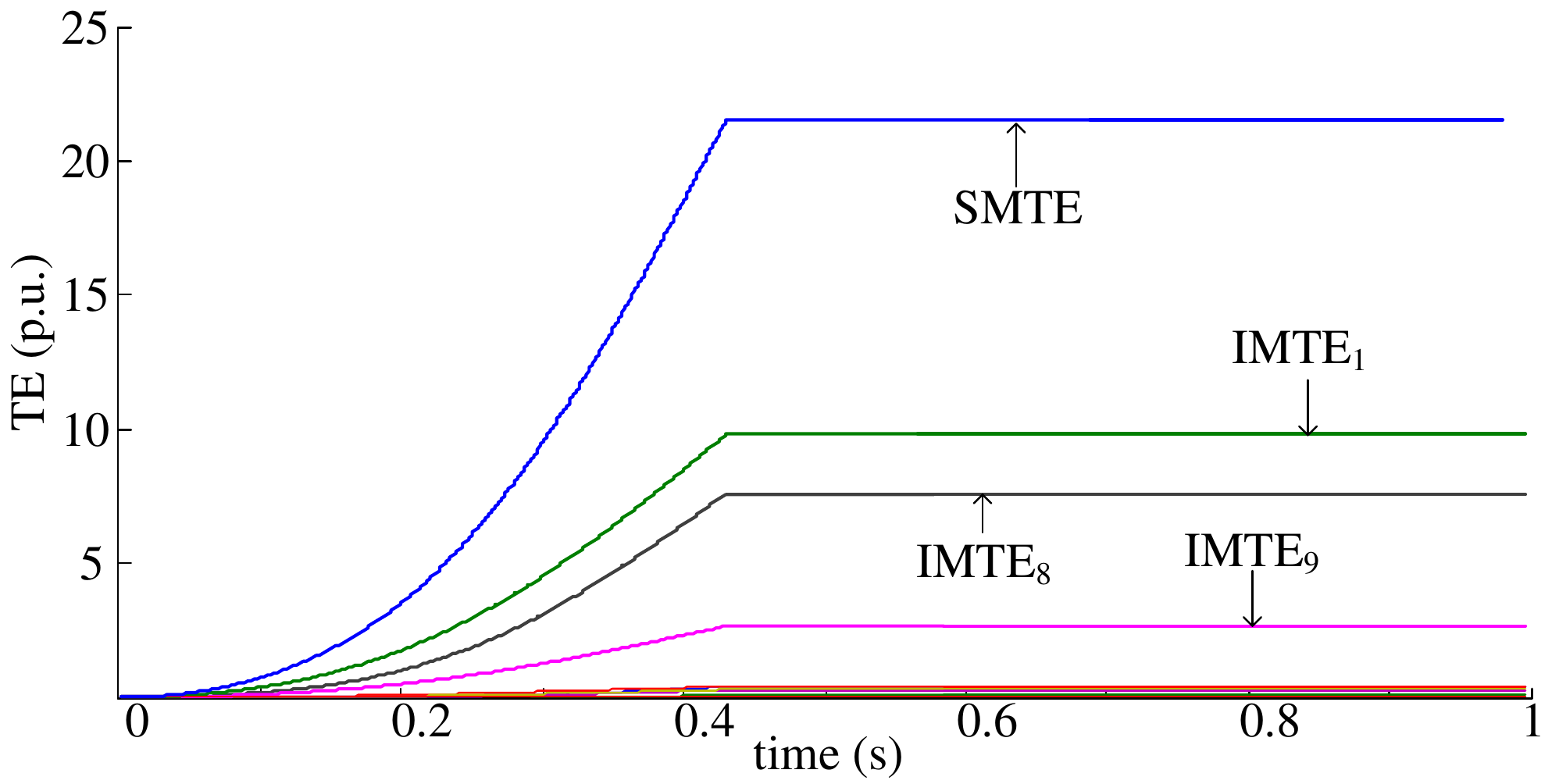}
  \caption{Conservativeness of SMTE [TS-1, bus-2, 0.430s].} 
  \label{fig3}  
\end{figure}

\noindent \textit{Explanation}: The conservativeness of the SMTE is ensured by the conserved IMTE of each machine in the system.
\\ \textit{Analysis}: The conservativeness of SMTE is demonstrated as below. The simulated original system trajectory is given in Ref. \cite{8}. Machines 8, 9 and 1 are critical machines in this case. SMTE and the IMTE of each machine are shown in Fig. \ref{fig3}.
From the figure, SMTE keeps conserved because each IMTE keeps conserved.

\subsection{THE OCCURRENCE OF THE SMPP (CHARACTERISTIC-II)} \label{section_IIIB}
\noindent \textit{Explanation}: The occurrence of SMPP is caused by the ``energy superimposition” of the maximum IMPE of each critical machine.
\\
\textit{Analysis}: From individual-machine perspective \cite{8}, the characteristic of the IMPE can be given as
\vspace*{0.5em}
\\ (a) The IMPE of a critical machine (no matter it maintains stable or goes unstable) fluctuates severely after fault clearing.
\\ (b) The IMPE of a critical machine reaches maximum at its corresponding DSP or DLP.
\\ (b) The IMPE of a non-critical machine varies slightly after fault clearing.
\par Based on (a) to (c), because the maximum IMPE of each critical machine is much higher than that of a non-critical machine, the ``peak” of the IMPE of each critical machine shows severe effect to the SMPE after superimposition. That is, the SMPE is certain to show a ``peak” with the occurrence of the SMPP after superimposition.
\par
\begin{figure}[H]
  \centering
  \includegraphics[width=0.45\textwidth,center]{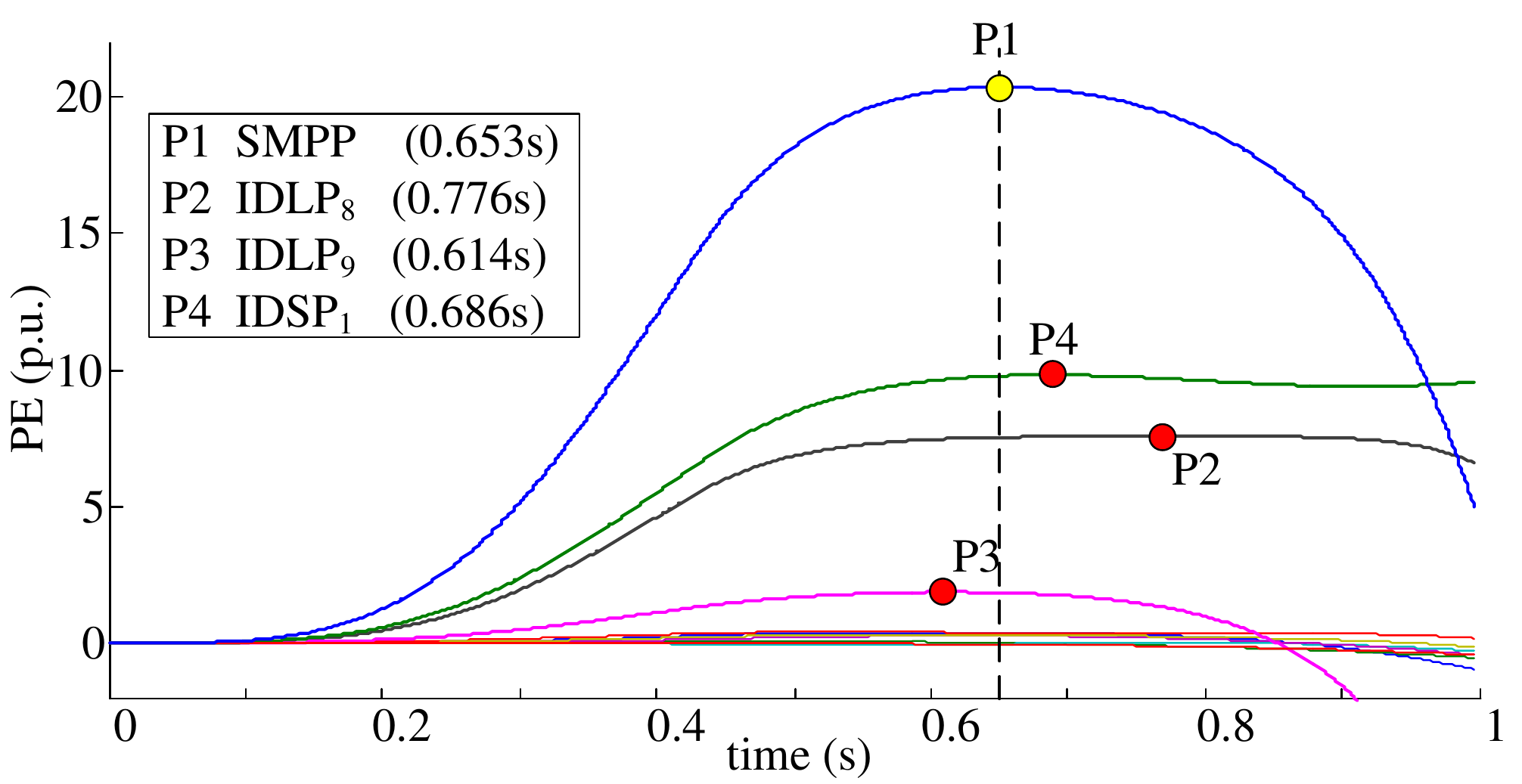}
  \caption{Occurrence of SMPP [TS-1, bus-2, 0.430s].} 
  \label{fig4}  
\end{figure}

 The occurrence of the SMPP along time horizon is shown in Fig. \ref{fig4}.
\vspace*{-0.5em}
From Fig. \ref{fig4}, The IMPEs of Machines 8, 9 and 1 are significantly higher than those of non-critical machines, and each IMPE shows a clear ``increasing-decreasing” characteristic.
\par The IDSP or IDLP of each critical machine is shown as below.
\vspace*{0.5em}
\\ $IDLP_9$ \textit{occurs (0.614 s)}: Machine 9 goes unstable. $\text{IMPE}_9$ reaches the maximum (1.863 p.u.).
\\ $IDSP_1$ \textit{occurs (0.686 s)}: Machine 1 maintains stable. $\text{IMPE}_1$ reaches the maximum (9.809 p.u.).
\\ $IDLP_8$ \textit{occurs (0.776 s)}: Machine 8 goes unstable. $\text{IMPE}_8$ reaches the maximum (7.571 p.u.).
\vspace*{0.5em}
\par Because the IMPE of each critical machine shows a ``peak” at IDSP or IDLP along time horizon, SMPE is certain to reach a maximum along time horizon after superimposition. In particular, SMPP occurs at 0.653 s and it just lies among the IMPPs of the three critical machines, as in Fig. \ref{fig4}.

\subsection{NEGATIVE INFINITE OF SMPE (CHARACTERISTIC-III)} \label{section_IIIC}
\noindent\textit{Explanation}: The negative infinite of the SMPE for an unstable system is caused by the ``energy superimposition” of each negative infinite IMPE of the unstable critical machine.
\\ \textit{Analysis}: With the increase of the simulation time, the characteristics of the IMPE are given as \cite{8}:
\\ (a) The IMPE of an unstable critical machine goes infinite along time horizon after it reaches maximum.
\\ (b) The IMPE of a stable critical machine is bounded along time horizon although it fluctuates severely.
\\ (c) The IMPE of a non-critical machine is bounded along time horizon because it varies slightly.
\par Following (a) to (c), because the IMPE of the unstable critical machine goes negative infinite along time horizon, the SMPE is certain to go negative infinite after energy superimposition.
\par
\begin{figure}[H]
  \centering
  \includegraphics[width=0.45\textwidth,center]{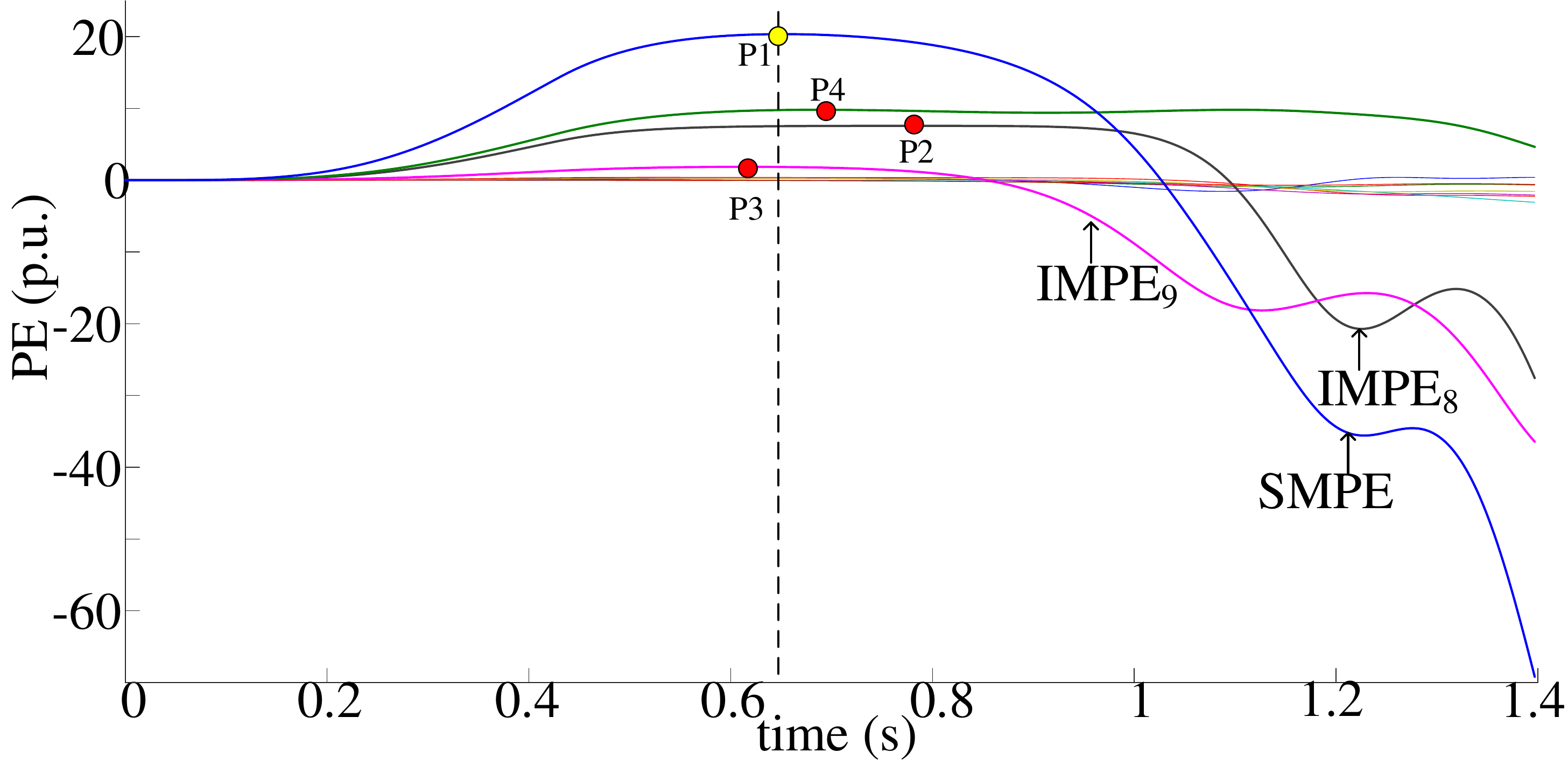}
  \caption{Negative infinity of SMPE along time horizon [TS-1, bus-2, 0.430s].} 
  \label{fig5}  
\end{figure}

 The variance of SMPE and IMPE along time horizon is shown in Fig. \ref{fig5}. The simulation time for the case in Fig. \ref{fig4} is extended to 1.400 s. The IMPE of each machine at 1.400 s is given in Table \ref{table1}.
\vspace*{-1em}
\begin{table}[H]
  \caption{IMPE of each machine at 1.400 s}
  \begin{tabular}{cc|cc}
  \hline
  Machine No. & IMPE (p.u.) & Machine No. & IMPE (p.u.)   \\ \hline
  10          & 0.397       & 7           & -1.558        \\
  3           & -0.669      & 8           & {\ul -27.579} \\
  4           & -2.242      & 9           & {\ul -36.445} \\
  5           & -3.080      & 1           & 4.643         \\
  6           & -2.075      & 2           & -0.595        \\ \hline
  \end{tabular}
  \label{table1}
\end{table}
\vspace*{-0.5em}
From Fig. \ref{fig5} and Table \ref{table1}, at 1.400 s, among the IMPEs of all machines in the system, $\text{IMPE}_9$ and $\text{IMPE}_8$ go negative
infinite along time horizon. Therefore, it is clear that the negative infinity of $\text{IMPE}_9$ and $\text{IMPE}_8$ finally cause the negative infinity of 
SMPE along time horizon after superimposition.

\subsection{ALWAYS EXISTANCE OF THE RESIDUAL SMKE  (CHARACTERISTIC-IV)} \label{section_IIID}
\begin{figure}[H]
  \centering
  \includegraphics[width=0.45\textwidth,center]{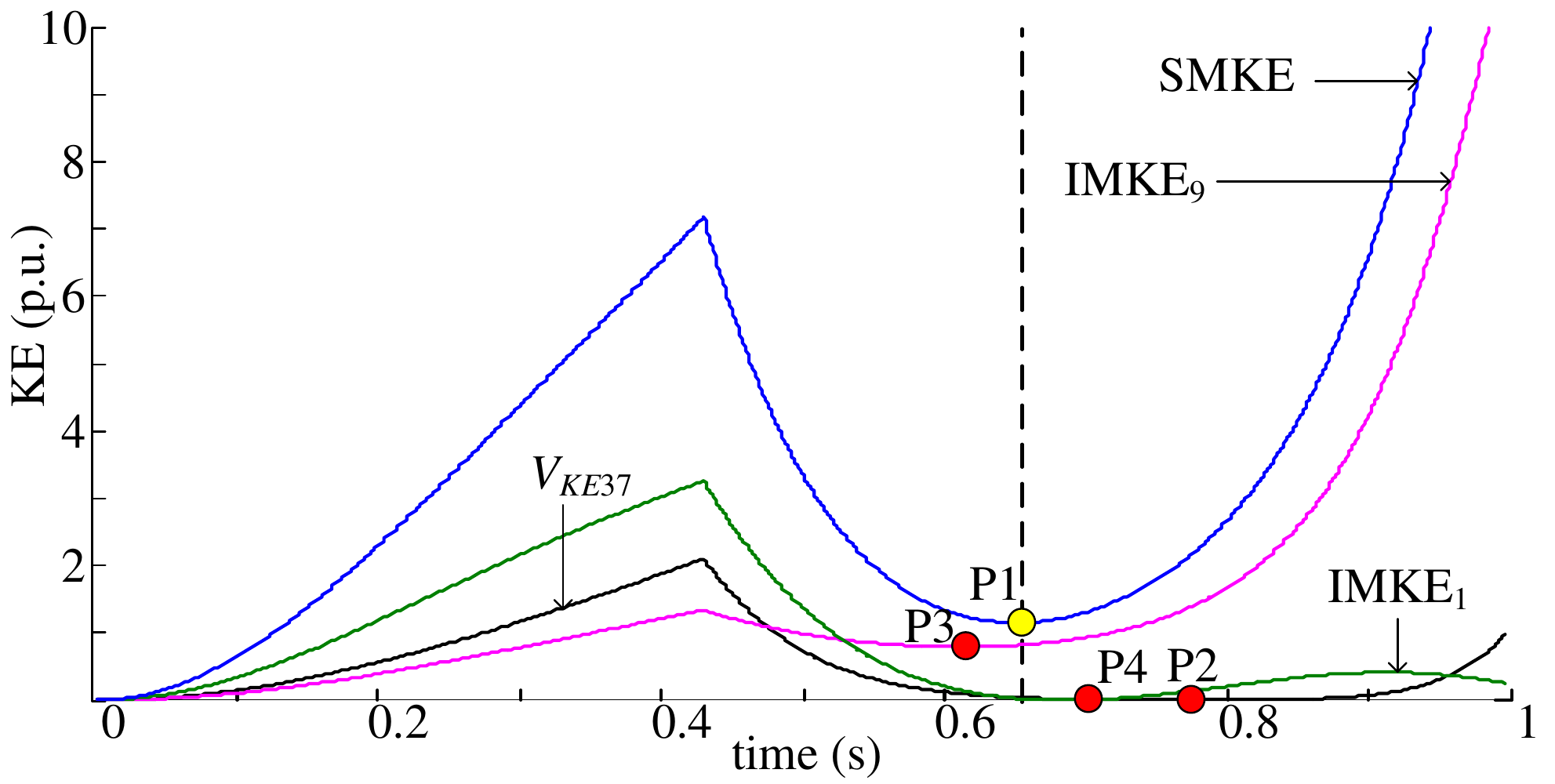}
  \caption{Always existence of the residual SMKE [TS-1, bus-2, 0.430 s].} 
  \label{fig6}  
\end{figure}

\noindent \textit{Explanation}: The always existence of the residual SMKE is caused by the ``energy superimposition” of IMKEs of all machines in the system at SMPP, as in Eq. (\ref{equ4}).
\\ \textit{Analysis}: The residual SMKE is shown in Fig. \ref{fig6}. The IMKE of each machine at SMPP is shown in Table \ref{table2}.
\vspace*{-1em}
\begin{table}[H]
  \small
  \centering
  \caption{IMKE of each machine at SMPP [TS-1, bus-2, 0.430 s].}
  \begin{tabular}{cc|cc}
  \hline
  \begin{tabular}[c]{@{}c@{}}Machine \\ No.\end{tabular} & \begin{tabular}[c]{@{}c@{}}IMKE at SMPP\\ (p.u.)\end{tabular} & \begin{tabular}[c]{@{}c@{}}Machine \\ No.\end{tabular} & \begin{tabular}[c]{@{}c@{}}IMKE at SMPP\\ (p.u.)\end{tabular} \\ \hline
  10                                                     & 0.032                                                         & 7                                                      & 0.007                                                         \\
  3                                                      & 0.080                                                         & 8                                                      & 0.043                                                         \\
  4                                                      & 0.025                                                         & 9                                                      & 0.818                                                         \\
  5                                                      & 0.051                                                         & 1                                                      & 0.029                                                         \\
  6                                                      & 0.002                                                         & 2                                                      & 0.064                                                         \\ \hline
  \end{tabular}
  \label{table2}
\end{table}
\vspace*{-0.5em}
From Fig. \ref{fig6} and Table \ref{table2}, at the moment that SMPP occurs, the IMKE of each machine in the system is positive. Therefore, the superimposed residual SMKE is always positive (1.151 p.u.).
\par From the analysis above, all the transient characteristics of the superimposed machine are essentially caused by the ``energy superimposition”.
However, this energy superimposition causes the machine to become a pseudo machine without equation of motion. Against this background, the superimposed machine violates all the machine paradigms and it shows problems in TSA. Detailed analysis is given in the following section.

\section{SUPERIMPOSED MACHINE BASED ORIGINAL SYSTEM STABILITY} \label{section_IV}
\subsection{VIOLATIONS OF THE MACHINE PARADIGMS} \label{section_IVA}
Following the analysis in Ref. \cite{1}, the equation of motion of the machine plays the ``dominant” role in the machine paradigms. This can be further expressed as
\\
\par \textit{The trajectory variance is modeled through the two-machine system}.
\par \textit{The transient energy conversion is also defined based on the two-machine system}.
\\
\par In other words, the entire machine paradigms are established based on the equation of motion of the machine.
\par However, following the analysis in Sections \ref{section_II} and \ref{section_III}, because of the global monitoring, the ``separation” of an objective machine with respect to the COI-SYS is unable to be established.
Then, the two-machine system is unable to be modeled. Under this background, the superimposed machine becomes the ``pseudo” machine without equation of motion.
Further, also due to the missing of the equation of motion, the superimposed machine transient energy conversion cannot satisfy strict NEC as residual SMKE always exists (Characteristic-IV).
\par  The missing of equation of motion in the superimposed machine indicates that this pseudo machine completely violates all the machine paradigms. These violations are given as below.
\\ \textit{Violation of the trajectory paradigm}: From Section \ref{section_IIA}, under global monitoring, the separation of the objective machine in the COI-SYS reference cannot be found. Therefore, this global monitoring violates the trajectory paradigm.
\\ \textit{Violation of the modeling paradigm}: From Section \ref{section_IIB}, because of the global monitoring, the two-machine system is unable to be modeled. Under this background, the superimposed machine does not have equation of motion. Therefore, the superimposed machine violates the modeling paradigm.
\\ \textit{Violation of the energy paradigm}: From Section \ref{section_IIC}, The SMTE is defined through the superimposition of the IMTEs of all machines in the system. This superimposition directly causes the failure of NEC because residual SMKE always exists at SMPP. Therefore, this superimposed-machine transient energy conversion violates the energy paradigm. The EAC can neither be found in the superimposed machine.
\par In brief, all the violations of the machine paradigms of the superimposed machine are fully caused by the ``energy superimposition” as analyzed in Section \ref{section_III}.
This is because the equation of motion of the superimposed machine is unable to be established through this energy superimposition. The failure of the two machine modeling further causes the pseudo NEC with the always existence of the residual SMKE.

\subsection{SUPERIMPOSED MACHINE STABILITY AND THE SYSTEM STABILITY} \label{section_IVB}
\noindent \textit{Superimposed machine stability}: The stability margin of the superimposed machine is denoted as
\begin{equation}
  \label{equ4}
  \eta_{\mathrm{SM}}=\frac{V_{\mathrm{SM}\mbox{-}\mathrm{SYS}}^{\mathrm{cr}}-V_{\mathrm{SM}\mbox{-}\mathrm{SYS}}^{\mathrm{c}}}{V_{K E \mathrm{SM}\mbox{-}\mathrm{SYS}}}
\end{equation}
\par From Eq. (\ref{equ4}), the stability state of the superimposed machine is characterized through the sign of $\eta_{\mathrm{SM}}$: $\eta_{\mathrm{SM}}>0$ means that the machine is stable;
$\eta_{\mathrm{SM}}=0$ means that the machine is critical stable; and $\eta_{\mathrm{SM}}<0$ means that the machine becomes unstable; The stability margin, i.e., the ``severity” of the machine is measured through the absolute value of $\eta_{\mathrm{SM}}$.
\par In fact, different from the margin definition of the individual machine that strictly follows the machine paradigms \cite{2}, Eq. (\ref{equ4}) cannot satisfy NEC characteristic because the superimposed machine is pseudo.
\\ \textit{System stability}: Following the analysis as in Section \ref{section_II}, because superimposed machine is the ``one-and-only” machine in TSA, the system stability is identical to the machine stability. Therefore, the stability margin of the system is defined as
\begin{equation}
  \label{equ5}
  \eta_{\mathrm{sys}}=\eta_{\mathrm{SM}}
\end{equation}
\par  Eq. (\ref{equ5}) indicates that the stability and severity of the system will be obtained simultaneously in TSA. However, this system stability evaluation does not satisfy NEC because the superimposed machine is a pseudo machine whose residual SMKE always exists.

\subsection{PROBLEMS OF THE SUPERIMPOSED MACHINE IN TSA} \label{section_IVC}
From the analysis in Section \ref{section_II}, the superimposed-machine analysts focus on the transient characteristics of the one-and-only superimposed machine in the original system. However, because of the global monitoring, the superimposed machine becomes a pseudo machine without equation of motion. It only has pseudo NEC with the always existence of the residual SMKE.
\par
\begin{figure}[H]
  \centering
  \includegraphics[width=0.48\textwidth,center]{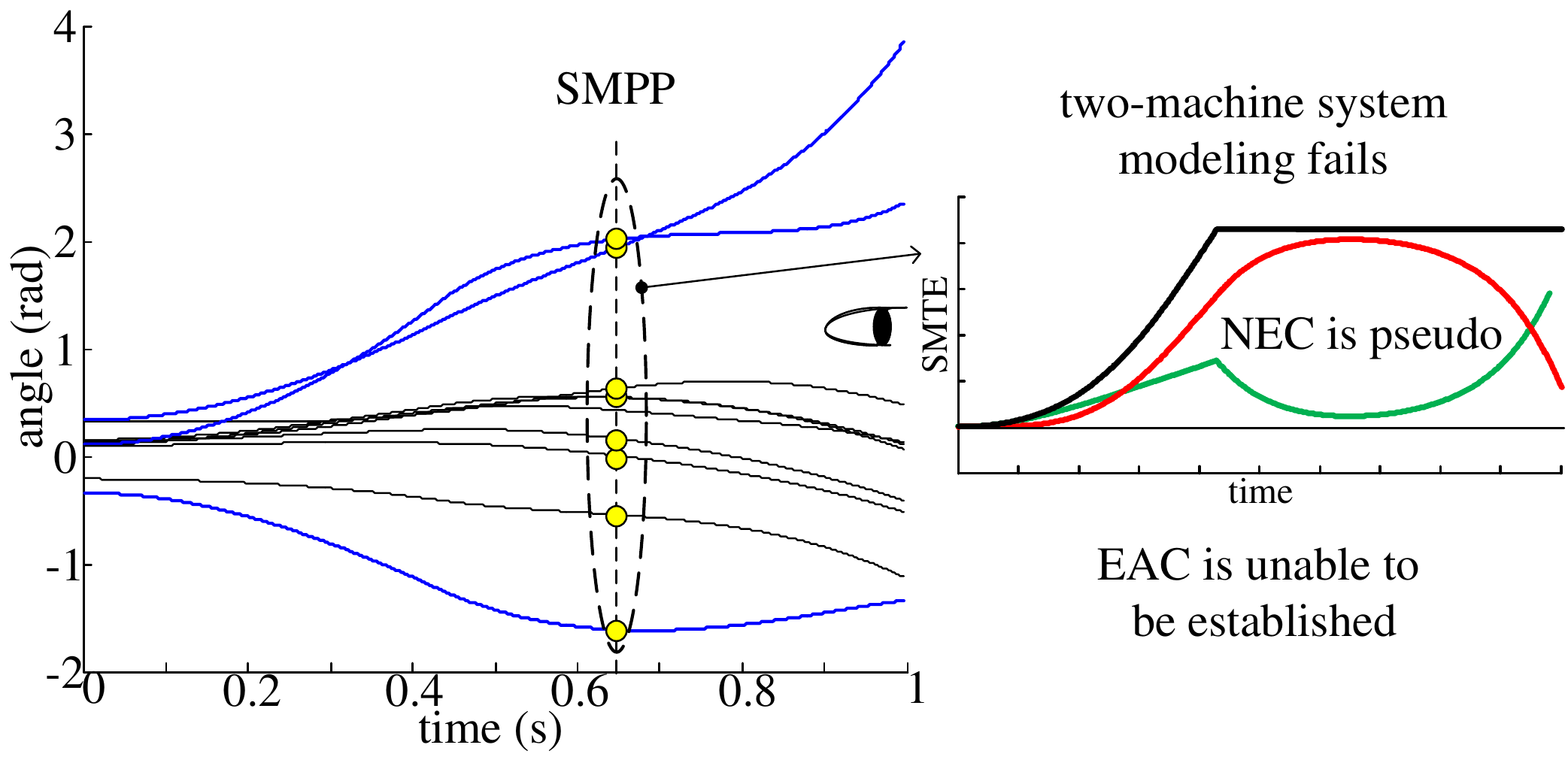}
  \caption{The use of the superimposed machine in the original system [TS-1, bus-2, 0430 s].} 
  \label{fig7}  
\end{figure}

  The use of the superimposed machine in the original system is shown in Fig. \ref{fig7}.
Frankly, the superimposed machine transient energy conversion is physically ``meaningless”. In particular, because the superimposed machine only has pseudo NEC without the equation of motion, the superimposed machine violates all the machine paradigms. These violations bring the following two problems
\\ \textit{Problem-I) The superimposed machine always have residual SMKE at SMPP}.
\\ \textit{Problem-II) The superimposed machine does not have corresponding trajectory}.
\vspace*{0.5em}
\par The two problems above can also be expressed in another form.
\\ \textit{Problem-I) The NEC fails in the superimposed machine}.
\\ \textit{Problem-II) The original system trajectory and the SMTE do not match}.
\vspace*{0.5em}
\par These two problems indicate that the superimposed machine shows the following two defects in TSA
\vspace*{0.5em}
\\ Stability-characterization defect: The stability of the original system trajectory is unable to be characterized precisely at SMPP (caused by Problem-I).
\\ Trajectory-depiction defect: The variance of the original system trajectory is unable to be depicted clearly through the SMPP (caused by Problem-II).
\vspace*{0.5em}
\par The two defects will be fully reflected in the definitions of the superimposed-machine based transient stability concepts. This will be analyzed in Section \ref{section_V}.\\\\

\section{DEFECTS WITH THE DEFINITIONS OF THE SUPERIMPOSED MACHINE BASED TRANSIENT STABILITY CONCEPTS} \label{section_V}
\subsection{DEFECT WITH SUPERIMPOSED MACHINE SWING} \label{section_VA}
\noindent \textit{Statement}: The trajectory-depiction defect is fully reflected in the definition of the superimposed-machine swing.
\\ \textit{Superimposed-machine perspective}: he system engineer monitors the original system trajectory in a global manner. The superimposed-machine swing is also defined as the SMPP. Further, because the superimposed-machine is the ``one-and-only” machine in the system, the superimposed-machine swing is also seen as the one-and-only swing of the original system.
\par
\begin{figure}[H]
  \centering
  \includegraphics[width=0.37\textwidth,center]{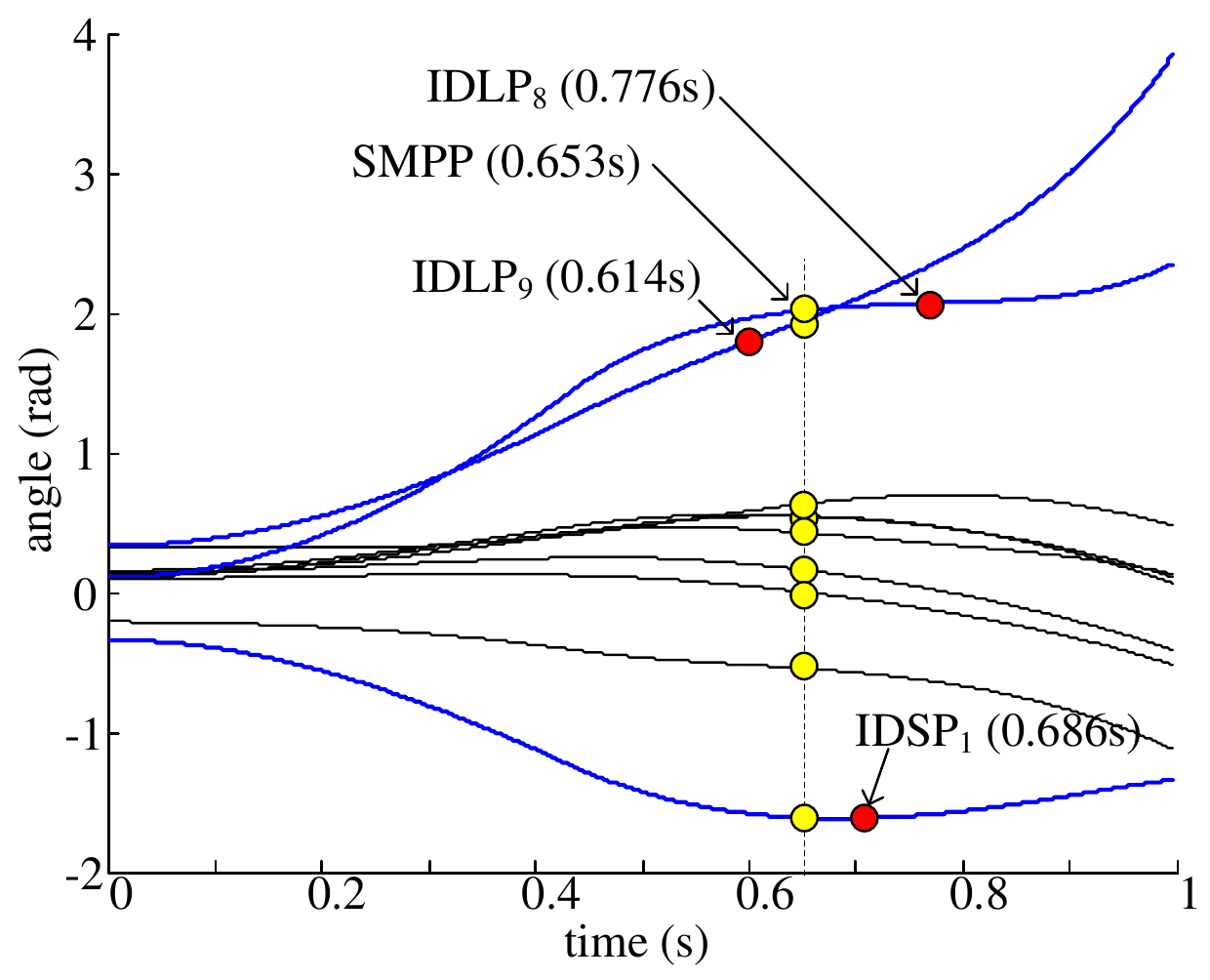}
  \caption{Unstable system trajectory [TS-1, bus-2, 0.430 s].} 
  \label{fig8}  
\end{figure}

 Unfortunately, this global swing definition will show problems in TSA, because the IMTR variance of each critical machine in the original system becomes confusing at SMPP \cite{2}.
\\ \textit{Example}: The simulation case is given to demonstrate the problem of the superimposed-machine swing. The unstable system trajectory is shown in Fig. \ref{fig8}.
Machines 8, 9 and 1 are unstable critical machines in this case.
\vspace*{-0.5em}
In Fig. \ref{fig8}, from superimposed-machine perspective, the system is defined as first swing unstable at SMPP that occurs at 0.653s. However, the IMTR variance of each critical machine in stable original system are not clear at the moment. In particular, Machine 9 already goes first-swing unstable for a while, Machine 1 is accelerating in its second swing, and Machine 8 is decelerating in its first swing.
Therefore, the ``trajectory separation” of the unstable original system cannot be depicted clearly at SMPP.

\subsection{DEFECT WITH CRITICAL STABILITY OF THE SYSTEM} \label{section_VB}
\noindent \textit{Statement}: Both the stability-characterization defect and the trajectory-depiction defect are reflected in the definition of the critical stability of the original system.
\\ \textit{Superimposed-machine perspective}: The critical stability state of the original system is completely decided by the critical stability of the superimposed machine. Based on this, SMPP when the system maintains critical stable is named the ``superimposed machine critical transient energy point” (SCTP). SCTP was also believed to be the critical transient energy point of the system. 
\\ \textit{Example}: The simulation case is given to demonstrate the definitions of the critical stable original system trajectory through SCTP. The critical stable and the critically unstable original system trajectories are shown in Figs. \ref{fig9} (a) and (b), respectively.
The SCTP is also shown in Fig. \ref{fig9a}. Demonstration about the superimposition in GTE when system maintains critical stable is shown in Fig. \ref{fig10}. The residual SMKE in both critical-stable case and critical-unstable case is shown in Table \ref{table3}.
\begin{figure} [H]
  \centering 
  \subfigure[]{%
  \label{fig9a}
    \includegraphics[width=0.4\textwidth]{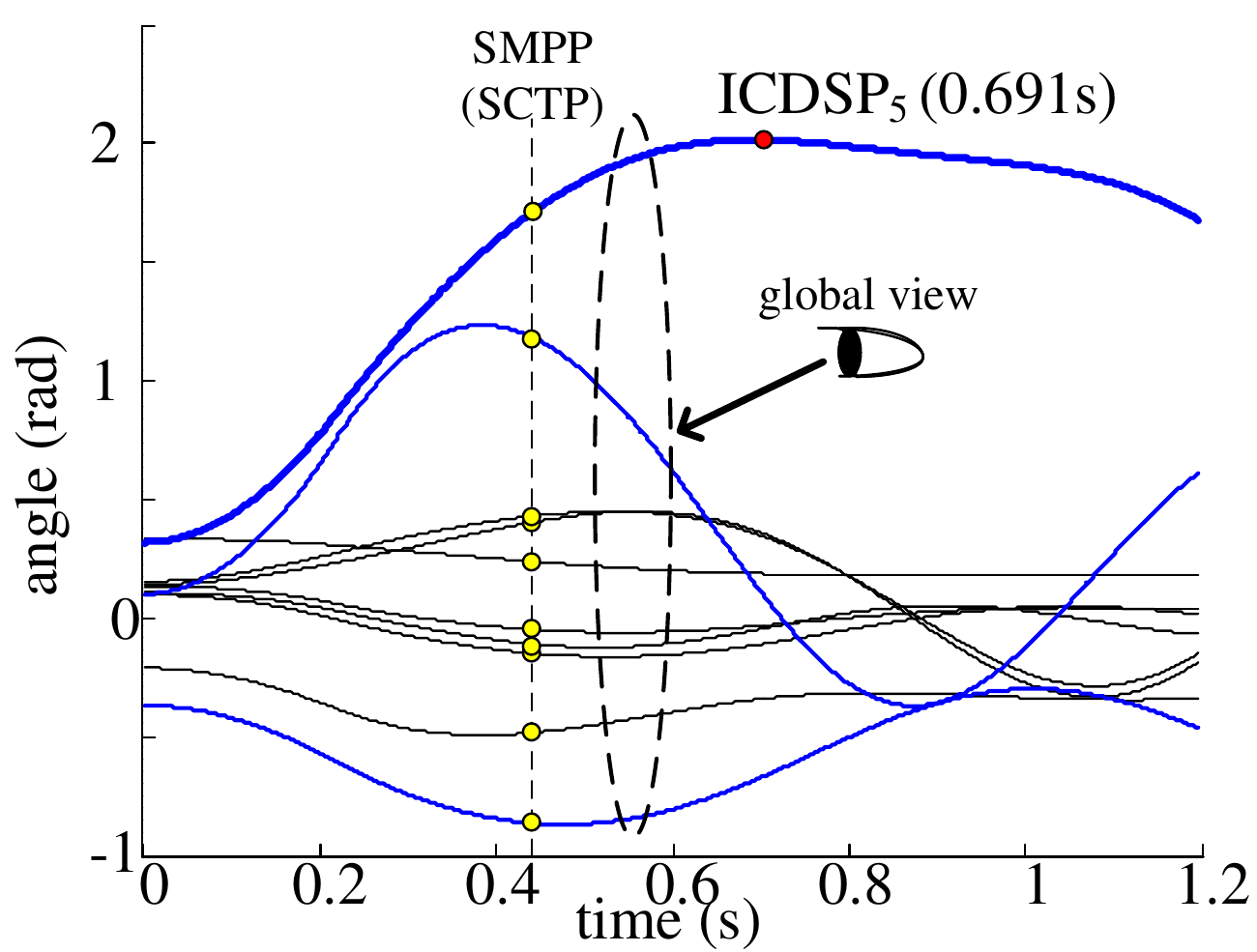}}%
\end{figure} 
\vspace*{-2em}
\addtocounter{figure}{-1}       
\begin{figure} [H]
  \addtocounter{figure}{1}      
  \centering 
  \subfigure[]{%
    \label{fig9b}
    \includegraphics[width=0.4\textwidth]{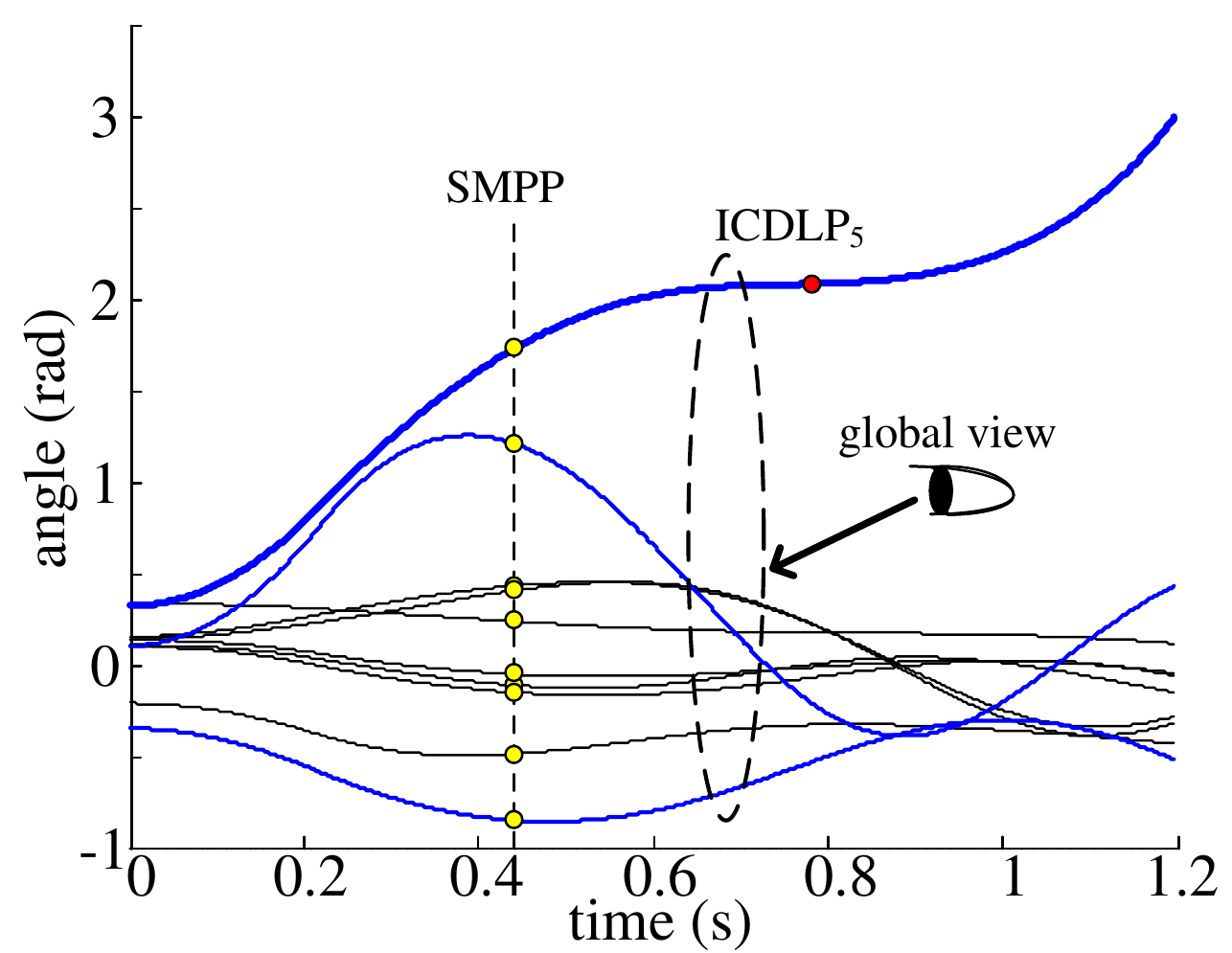}}%
  \caption{System trajectory. (a) Critical stable case [TS-1, bus-19, 0.215s]. (b) Critical unstable case [TS-1, bus-19, 0.216s].}%
  \label{fig9}
\end{figure}

\begin{figure}[H]
  \centering
  \includegraphics[width=0.37\textwidth,center]{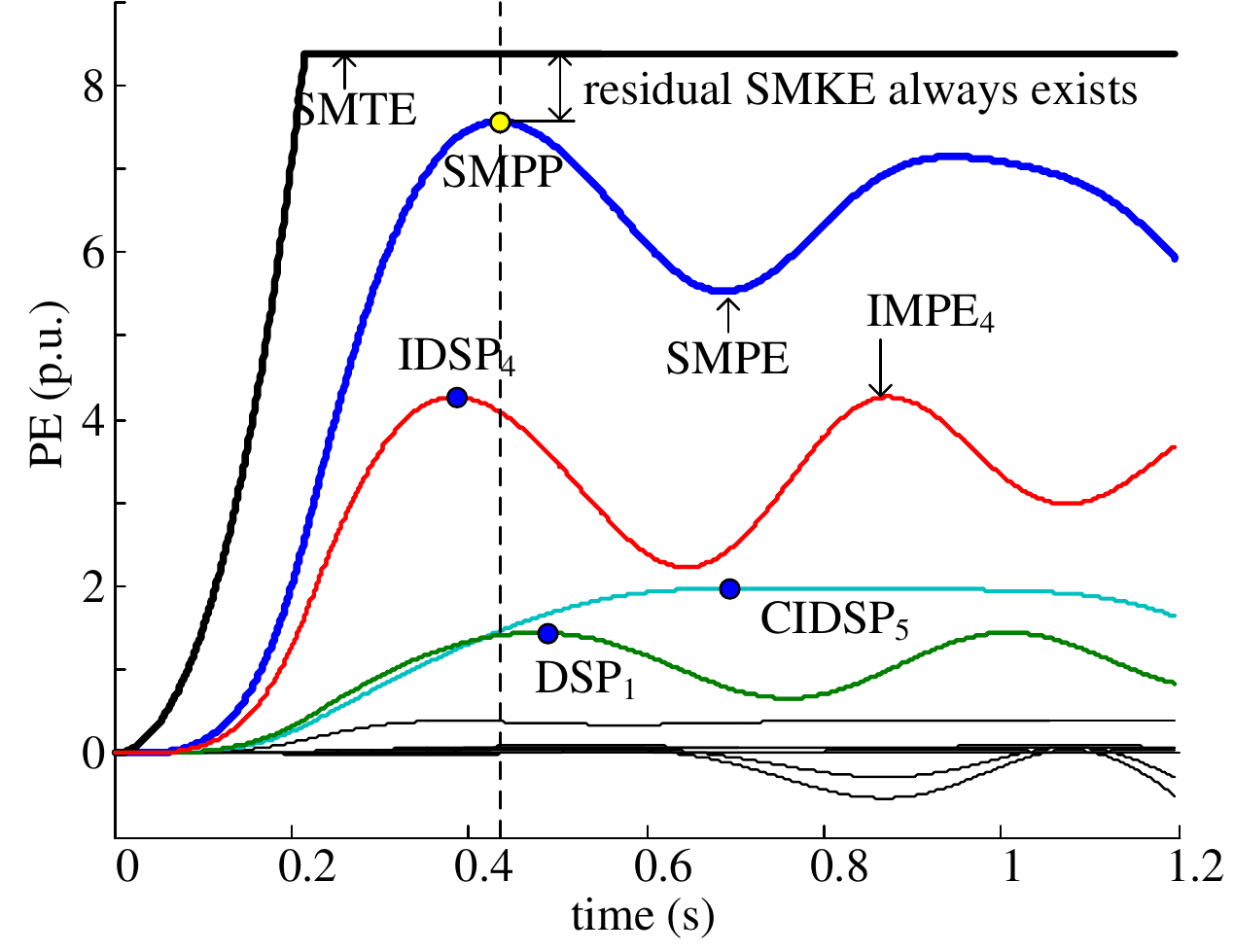}
  \caption{Superimposed effect in the GTE  [TS-1, bus-19, 0.215 s].} 
  \label{fig10}  
\end{figure}
\begin{table}[H]
  \footnotesize
  \centering
  \caption{Residual SMKE at the critical stable case}
  \begin{tabular}{@{}cccc@{}}
  \toprule
  System state      & \begin{tabular}[c]{@{}c@{}}SMKE at fault\\ clearing point\\ (p.u.)\end{tabular} & \begin{tabular}[c]{@{}c@{}}GPE at GMPP\\ (p.u.)\end{tabular} & \begin{tabular}[c]{@{}c@{}}Residual GKE\\ (p.u.)\end{tabular} \\ \midrule
  Critical stable   & 8.3731                                                                          & 7.5578                                                       & 0.8153                                                        \\
  Critical unstable & 8.5580                                                                          & 7.7223                                                       & 0.8357                                                        \\ \bottomrule
  \end{tabular}
  \label{table3}
\end{table}
\vspace*{-0.5em}
From Fig. \ref{fig10} and Table \ref{table3}, the analysis of the critical stability of the system is given below.
\\ Global monitoring: The critical stability of the system is decided by the critical stability of the superimposed machine.
Superimposed-machine transient energy conversion: The residual SMKE always exists no matter the system maintains critical stable or becomes critical unstable.
\par From analysis above, the pseudo superimposed shows two defects in the definitions of the critical stability of the system.
\\ (i) The stability state of the superimposed machine from the critical stability to the critical instability cannot be characterized precisely, because residual SMKE always exists (caused by Problem-I), as in Fig. \ref{fig10} and Table \ref{table3}.
\\ (ii) The trajectory variance of the original system trajectory from the critical stability to the critical instability cannot depicted clearly through the change of SMPP (caused by Problem-II), as in Fig. \ref{fig9}.
\par (i) and (ii) are fully caused by the complete violations of the machine paradigms in the superimposed machine.
\par
\begin{figure}[H]
  \centering
  \includegraphics[width=0.45\textwidth,center]{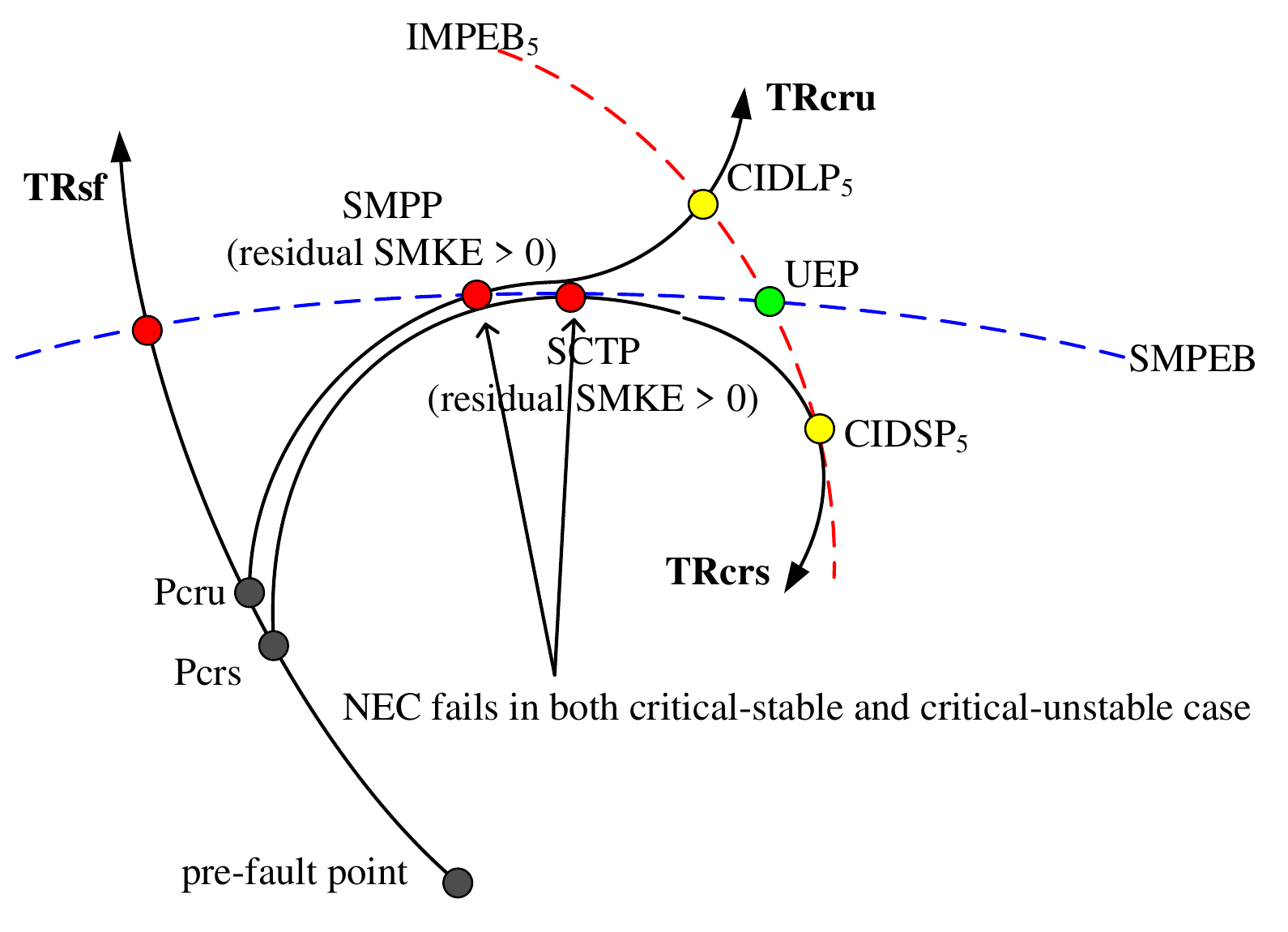}
  \caption{Demonstration of SCTP [TS-1, bus-19].} 
  \label{fig11} 
\end{figure}

 Demonstration of the critical stability of the superimposed machine from superimposed-machine perspective is shown in Fig. \ref{fig11}. The individual-machine expression is already given in Ref. \cite{2}.
\vspace*{-0.5em}
From Fig. \ref{fig11}, the SCTP occurs earlier than the $\text{IDSP}_{\text{MDM}}$ in this case. the crucial trajectory inflection and transient energy conversion of the critical stable system cannot be depicted precisely through SCTP.

\subsection{DEFECT WITH THE SMPES MODELING}   \label{section_VC}
\begin{figure}[H]
  \centering
  \includegraphics[width=0.42\textwidth,center]{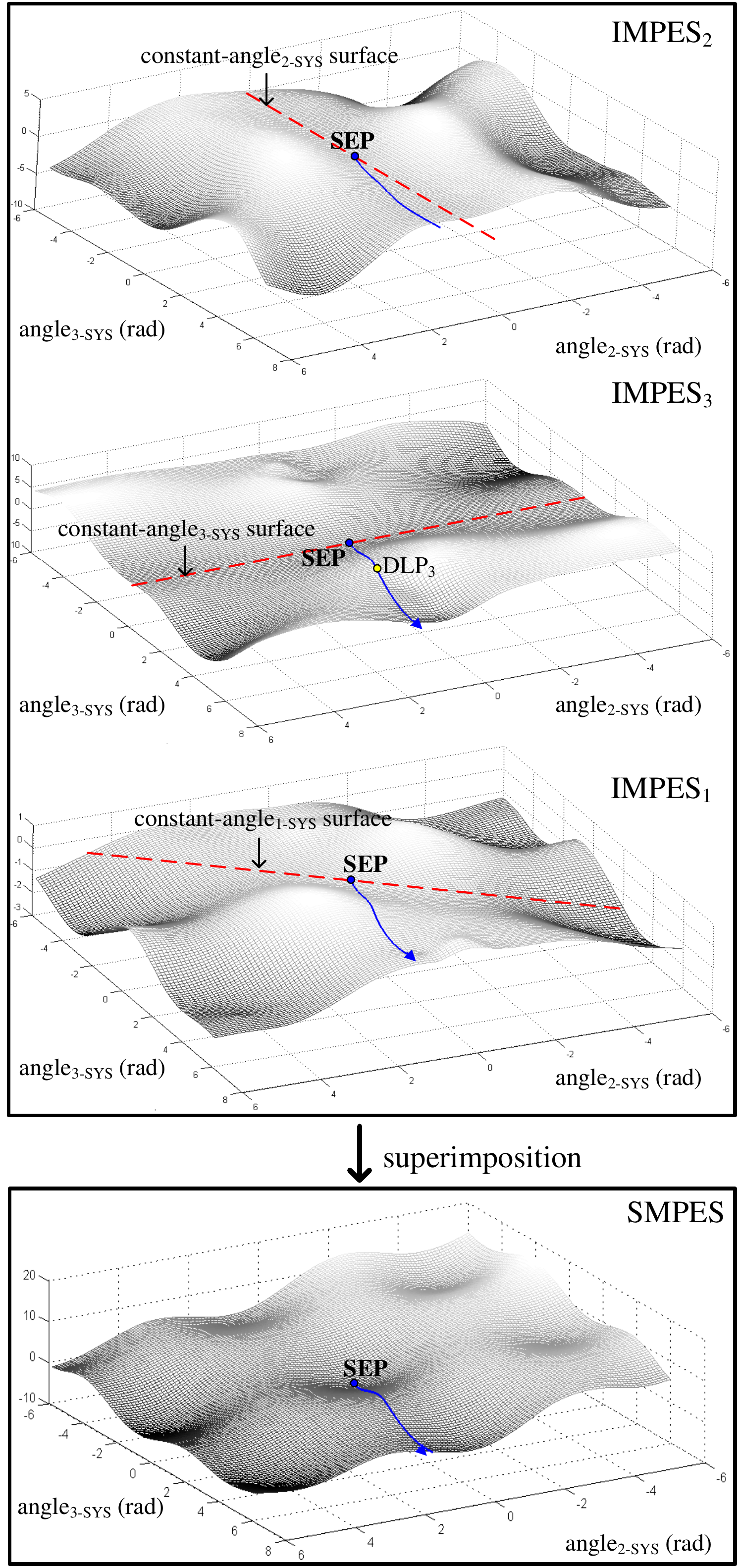}
  \caption{Formation of SMPES using TS-4 as the test bed.} 
  \label{fig12} 
\end{figure}

\begin{figure}[H]
  \centering
  \includegraphics[width=0.45\textwidth,center]{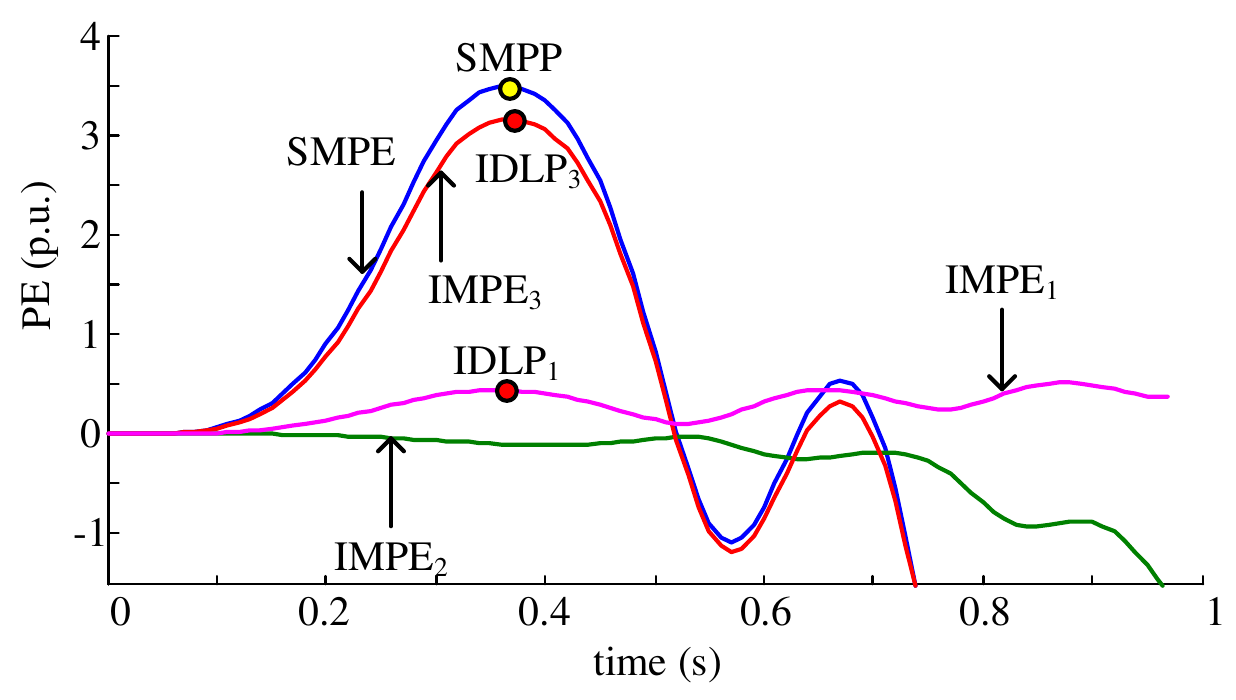}
  \caption{Occurrence of SMPP [TS-4, bus-3, 0.300 s].} 
  \label{fig13} 
\end{figure}

\noindent \textit{Statement}: The stability-characterization defect is fully reflected in the modeling of the superimposed-machine potential energy surface (SMPES).
\\ \textit{Superimposed-machine perspective}: The SMPES is modeled through the SMPE. Because the SMPE is the superimposition of the IMPEs of all machines in the system, the SMPES can also be seen as the superimposition of the IMPESs of all machines in the system.
\\ \textit{Example}: Demonstration about the superimposition of SMPES is shown in Fig. \ref{fig12}. The system trajectory is shown in Ref. \cite{9}. The SMPE is shown in Fig. \ref{fig13}.
\vspace*{0.5em}
\par From Fig. \ref{fig12}, the ``altitude” of the ball is the SMPE of the machine. Therefore, the energy ball rolling on the SMPES just reflects the superimposed machine transient energy conversion along original system trajectory (in the $\delta_{2\mbox{-}\mathrm{SYS}}$-$\delta_{3\mbox{-}\mathrm{SYS}}$ angle space), as in Fig. \ref{fig13}.
Note that this transient energy conversion does not satisfy NEC characteristic because the superimposed machine is pseudo.

\section{CASE STUDY} \label{section_VI}
\begin{figure}[H]
  \centering
  \includegraphics[width=0.4\textwidth,center]{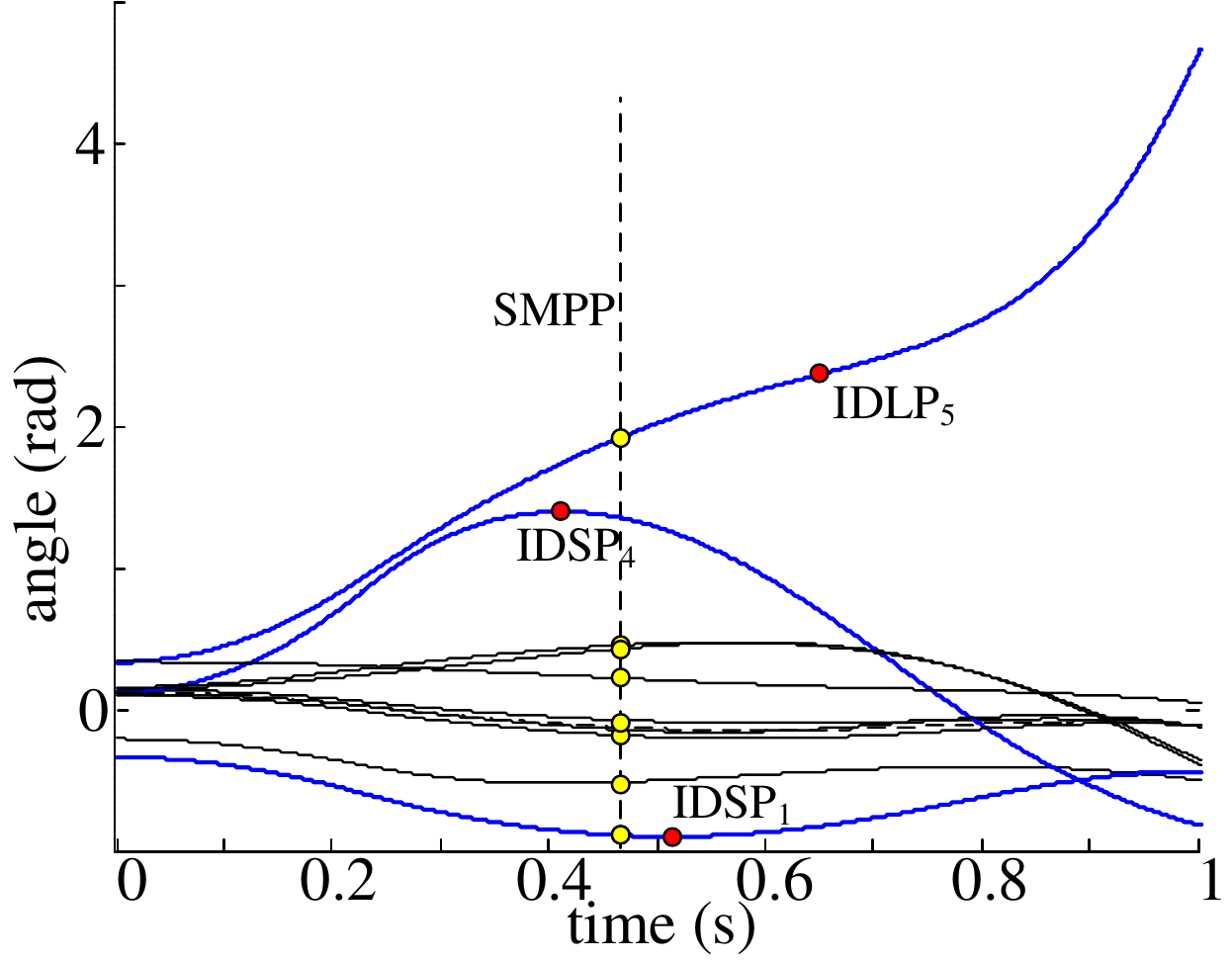}
  \caption{Original system trajectory [TS-1, bus-19, 0.230 s].} 
  \label{fig14} 
\end{figure}

\begin{figure}[H]
  \centering
  \includegraphics[width=0.4\textwidth,center]{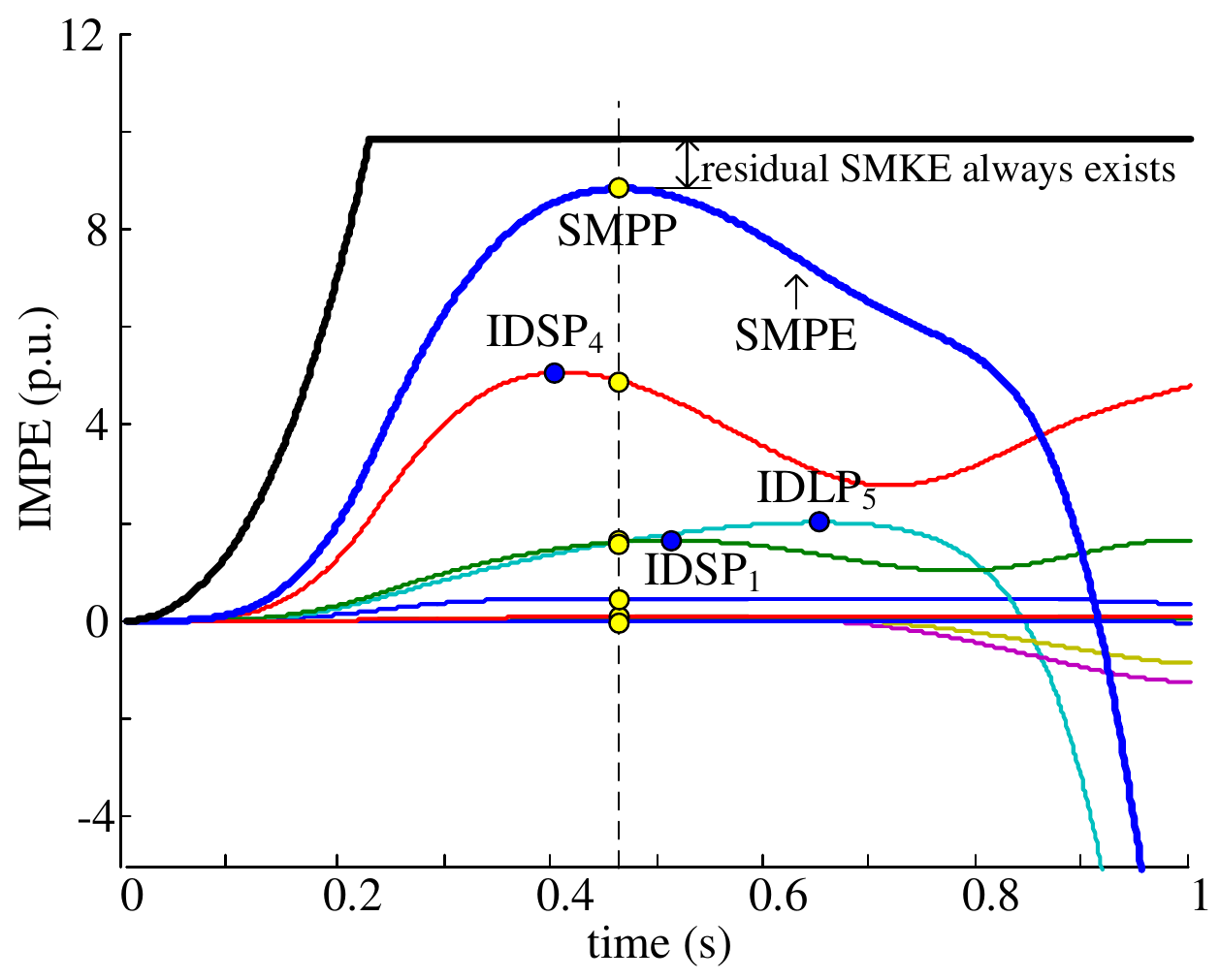}
  \caption{GPE along time horizon [TS-1, bus-19, 0.230 s].} 
  \label{fig15} 
\end{figure}

In this case the defects of the superimposed-machine in TSA will be demonstrated. The simulation case is the same with that in the Ref. \cite{2}. The system trajectory is shown in Fig. \ref{fig14}. The SMPE and IMPE along time horizon are shown in Fig. \ref{fig15}.
\vspace*{-2em}
\par The superimposed-machine based TSA is given as below
\\ \textit{Global trajectory monitoring}: Using COI-SYS as the motion reference, the system engineer monitors the original system trajectory in a global manner.
\\ \textit{Two-machine system modeling}: The two-machine system cannot be established.
\\ \textit{Superimposed machine stability evaluation}: The stability of the ``one-and-only” superimposed machine is seen as the representation of the stability of the entire original system. The stability of the superimposed machine is evaluated as below\
\vspace*{0.5em} 
\\\textit{SMPP occurs (0.468 s)}: The residual SMKE is 0.9907 p.u..
\vspace*{0.5em}
\par Following the analysis in Section \ref{section_III}, through the energy superimposition, NEC inside the machine completely fails because the residual SMKE  always exists at SMPP, as in Fig. \ref{fig15}. The stability of the system cannot be characterized because of the residual SMKE problem.
\par The IMKE of each machine at SMPP is shown in Table \ref{table4}.
\begin{table}[H]
  \small
  \centering
  \caption{IMKE of each machine at SMPP [TS-1, bus-19, 0.230s].}
  \begin{tabular}{cc|cc}
  \hline
  \begin{tabular}[c]{@{}c@{}}Machine \\ No.\end{tabular} & \begin{tabular}[c]{@{}c@{}}IMKE at GMPP\\ (p.u.)\end{tabular} & \begin{tabular}[c]{@{}c@{}}Machine \\ No.\end{tabular} & \begin{tabular}[c]{@{}c@{}}IMKE at GMPP\\ (p.u.)\end{tabular} \\ \hline
  10                                                     & 0.0097                                                         & 7                                                      & 0.0096                                                         \\
  3                                                      & 0.0083                                                         & 8                                                      & 0.0085                                                         \\
  4                                                      & 0.2432                                                         & 9                                                      & 0.0120                                                         \\
  5                                                      & {\ul 0.6317}                                                  & 1                                                      & 0.0215                                                         \\
  6                                                      & 0.0358                                                         & 2                                                      & 0.0104                                                         \\ \hline
  \end{tabular}
  \label{table4}
\end{table}
\vspace*{-0.5em}
From Fig. \ref{fig14}, the superimposed-machine swing also seems confusing at SMPP because each critical machine (especially Machine 5 that finally causes the system to go unstable) could not show a clear transient characteristic at SMPP.
In particular, Machine 4 is accelerating in its second swing, Machine 1 is decelerating in its first swing, while Machine 5 is still decelerating in the first swing.
\par
\begin{figure}[H]
  \centering
  \includegraphics[width=0.4\textwidth,center]{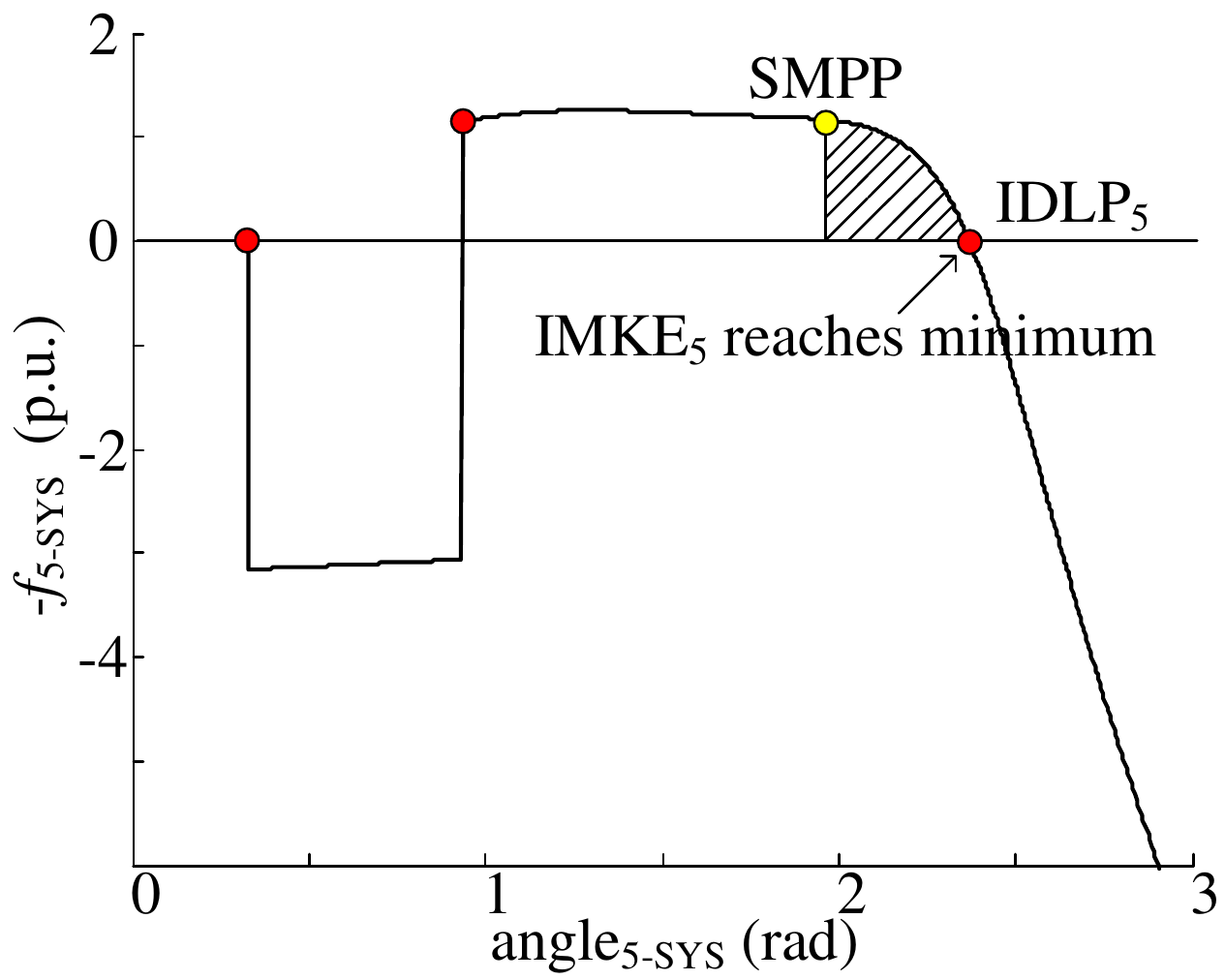}
  \caption{Kimbark curve of Machine 5.} 
  \label{fig16} 
\end{figure}

 The comparison between SMPP and IDLP is further demonstrated in the Kimbark curve of Machine 5, as in Fig. \ref{fig16}.
\vspace*{-0.5em}
From Fig. \ref{fig16}, at the moment that SMPP occurs, Machine 5 is decelerating along post-fault system trajectory. The machine is far from $\text{IDLP}_5$ where $\text{IMKE}_5$ reaches its minimum 0.2711 p.u..
Under this circumstance, Machine 5 cannot be characterized as unstable through its IMEAC at SMPP. The variance of the $\text{IMTR}_5$ at SMPP is also not quite clear, as in Figs. \ref{fig14}. \\\\

\section{VII.  THE THIRD DEFECT OF UEP} \label{section_VII}
\subsection{REVISIT OF THE DEFECT OF THE SUPERIMPOSED MACHINE BASED CRITCAL TRANSIENT ENERGY} \label{section_VIIA}
In Ref. \cite{9}, the first defect of UEP is exposed that the some UEPs physically do \textit{not} exist in an original multi-machine power system.
In Ref. \cite{10}, the second defect of UEP is exposed that UEP completely ignores the unique and different NEC characteristic inside each machine (UEP does not exist along actual post-fault system trajectory). In this paper, the third defect of UEP is exposed. That is, the UEP is used to compute the superimposed-machine based critical transient energy.
\par Following the analysis in Section \ref{section_VB}, the inherit problem of the superimposed-machine critical transient energy is that the residual SMKEs always exist when system maintains critical stable and becomes critical unstable.
That is, the critical stability of the system cannot be precisely depicted through the NEC characteristic of the SMTE (Problem-I).
\par TS-4 is used to demonstrate the inherit problem of the superimposed-machine critical transient energy. The fault is [TS-4, bus-7]. In this case UEP is computed as [2.104, 1.706, -0.787]. The simulation step is 0.01 s. The CCT is 0.23 s.
The critical stable and critical unstable system trajectories are shown in Figs. \ref{fig17} (a) and (b), respectively. The superimposed-machine transient energy conversion along actual critical unstable system trajectory is shown in Fig. \ref{fig18}.
The SMKE and SMPE at SMPP in the two cases are shown in Table \ref{table5}. In this small-scale test bed, all machines are MDMs. All the machines remain critical stable, and they become critical unstable with the slight increase of the disturbance. 
\begin{figure} [H]
  \centering 
  \subfigure[]{%
  \label{fig17a}
    \includegraphics[width=0.45\textwidth]{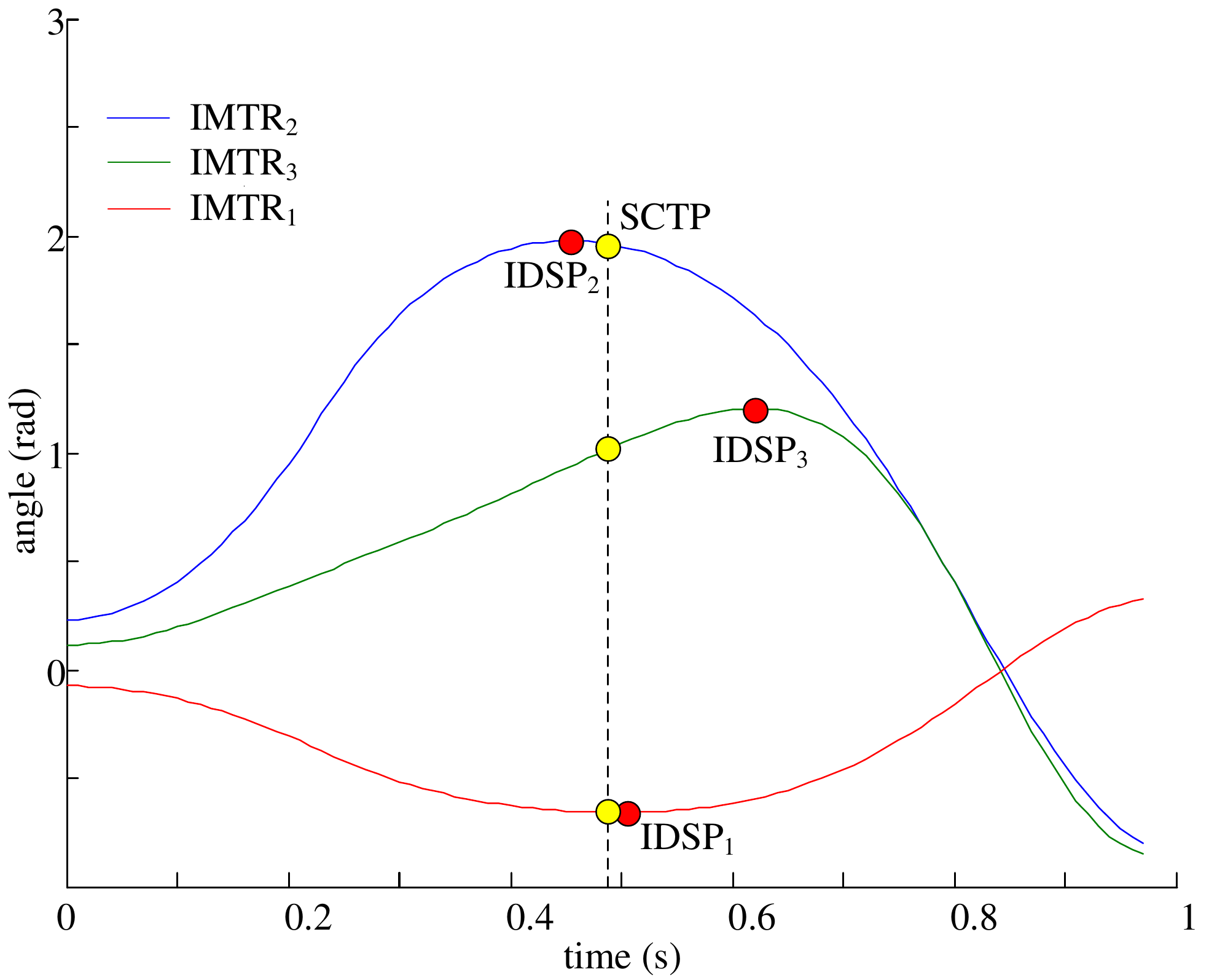}}%
\end{figure} 
\vspace*{-2em}
\addtocounter{figure}{-1}       
\begin{figure} [H]
  \addtocounter{figure}{1}      
  \centering 
  \subfigure[]{%
    \label{fig17b}
    \includegraphics[width=0.45\textwidth]{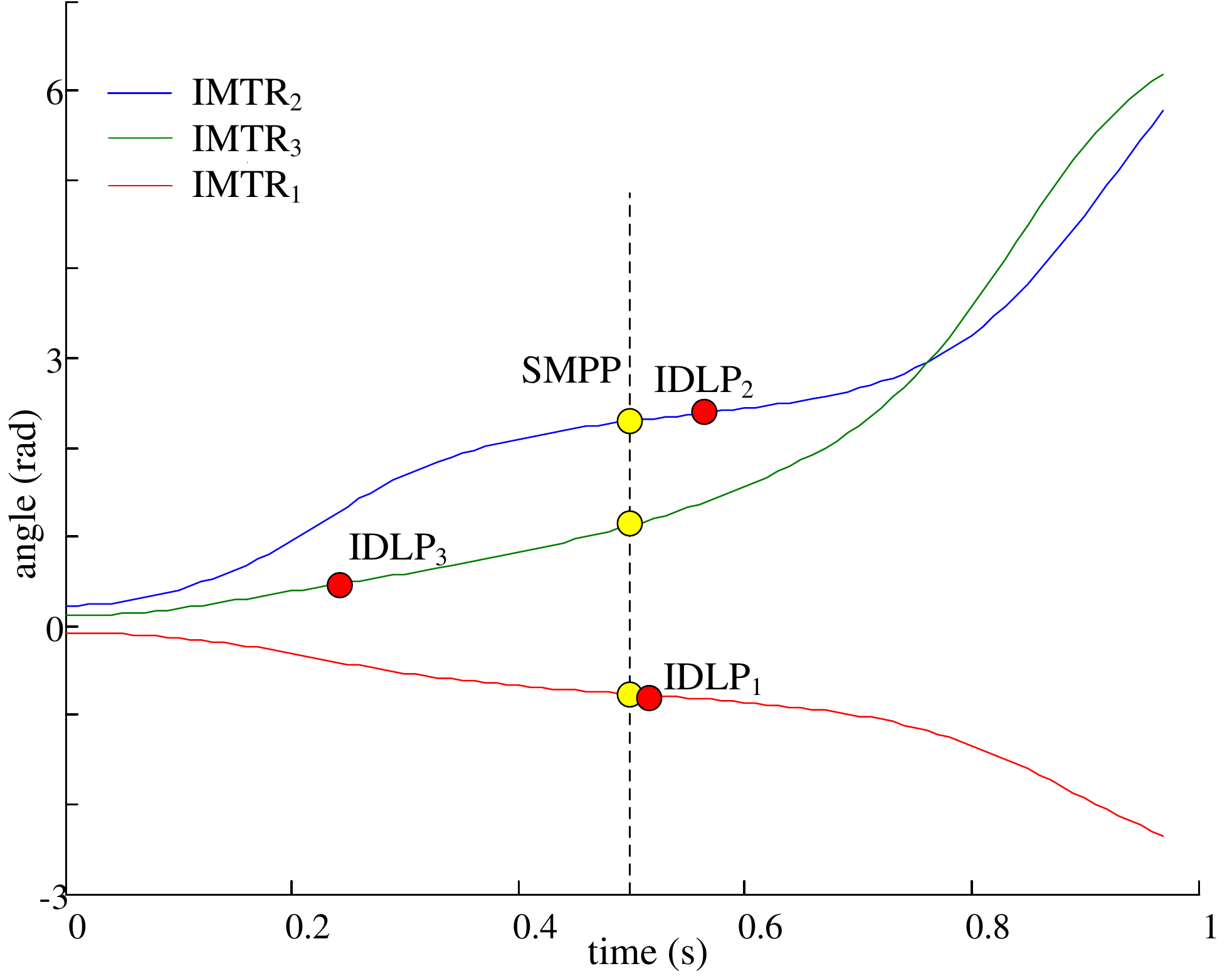}}%
  \caption{System trajectories. (a) Critical stable case [TS-4, bus-7, 0.23 s]. (b) Critical unstable case [TS-4, bus-7, 0.24 s].}%
  \label{fig17}
\end{figure}
\vspace*{-2em}
\begin{figure}[H]
  \centering
  \includegraphics[width=0.4\textwidth,center]{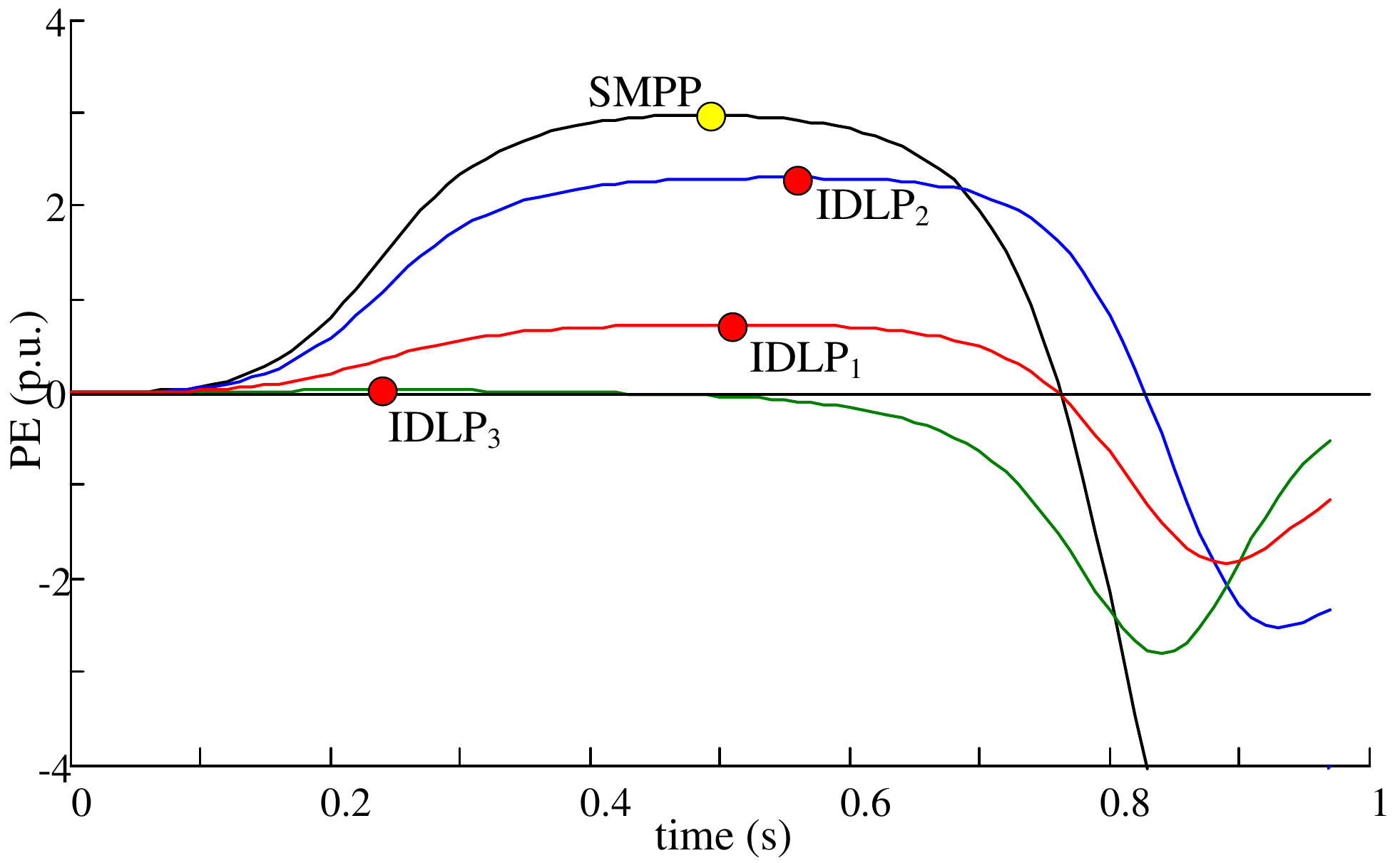}
  \caption{IMPE along time horizon [TS-4, bus-7, 0.24 s].} 
  \label{fig18} 
\end{figure}
\vspace*{-1em}
\begin{table}[H]
  \centering
  \caption{SMKE and SMPE at the SMPP in the two cases}
  \begin{tabular}{@{}cc|cc@{}}
  \toprule
  \multicolumn{2}{c|}{Critical stable case} & \multicolumn{2}{c}{Critical unstable case} \\ \midrule
  SMKE (p.u.)            & 0.0470           & SMKE (p.u.)             & 0.1774           \\
  SMPE (p.u.)            & 2.7932           & SMPE (p.u.)             & 2.9702           \\
  $\text{IMKE}_2$ (p.u.)           & 0.0028           & $\text{IMKE}_2$ (p.u.)            & 0.0398           \\
  $\text{IMKE}_3$ (p.u.)           & 0.0421           & $\text{IMKE}_3$ (p.u.)            & 0.0925           \\
  $\text{IMKE}_1$ (p.u.)           & 0.0021           & $\text{IMKE}_1$ (p.u.)            & 0.0451           \\ \bottomrule
  \end{tabular}
  \label{table5}
\end{table}
From Figs. \ref{fig17} and \ref{fig18}, following the analysis in Section \ref{section_VB}, the SCTP has problems in both stability characterization and trajectory depiction when the original system maintains critical unstable (Problem-I and Problem-II).
Therefore, using the superimposed machine based critical transient energy can be seen as a ``mistake” in TSA.

\subsection{LINEAR TRAJECTORY ASSUMPTION} \label{section_VIIB}
In order to solve the two problems of the SCTP, the UEP is used to compute the superimposed-machine critical transient energy that is already clarified to be a mistake in TSA.
\par Because UEP does not exist along actual post-fault system trajectory (the second defect of UEP as analyzed in Ref. \cite{10}), the involvement of the UEP brings the following assumption
\\
\par \textit{A fictional linear system trajectory should be created from SEP to UEP}.
\vspace*{0.5em}
\par The linear system trajectory is expressed as
\begin{equation}
  \label{equ6}
  \delta_{i\mbox{-}\mathrm{SYS}}^{(\mathrm {linear })}=\alpha\left(\delta_{i\mbox{-}\mathrm{SYS}}^{\mathrm{UEP}}-\delta_{i\mbox{-}\mathrm{SYS}}^{\mathrm{SEP}}\right)+\delta_{i\mbox{-}\mathrm{SYS}}^{\mathrm{SEP}}
\end{equation}
\par
\begin{figure}[H]
  \centering
  \includegraphics[width=0.4\textwidth,center]{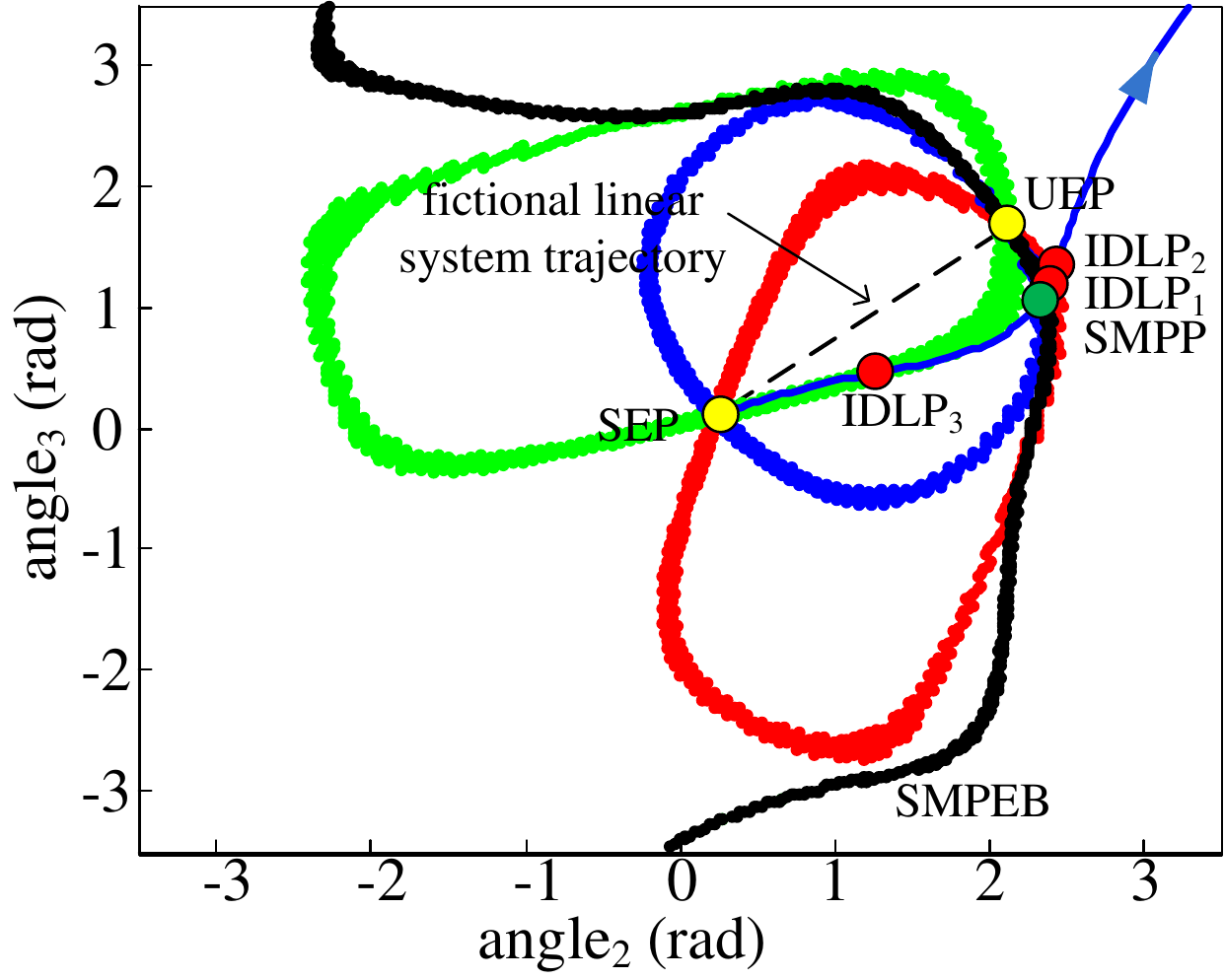}
  \caption{UEP in the angle space.} 
  \label{fig19} 
\end{figure}

 The UEP and the linear system trajectory are given as below. The actual simulated unstable system trajectory [TS-4, bus-7, 0.24 s] in the angle space is shown in Fig. \ref{fig19}.
UEP is computed as [2.104, 1.706, -0.787] in this case. The UEP and the linear system trajectory in the angle space is also shown in the figure. The SMPE and IMPE along time horizon is already shown in Fig. \ref{fig18}.
\vspace*{-1.5em}
\subsection{A FAR GREATER MISTAKE IN THE UEP-BASED SMPE} \label{section_VIIC}
From the analysis in Section \ref{section_VIIB}, mathematically, the UEP ([2.104, 1.706, -0.787]) is defined in an individual-machine form. However, actually, it is used to ``approximate” the superimposed-machine critical transient energy at SCTP. 
This approximation is based on the computation of a fictional SMPE at UEP along the linear system trajectory. This UEP based SMPE is also believed to be the critical transient energy along the ``fictional” linear system trajectory.
For the case in Fig. \ref{fig19}, along linear trajectory, the UEP based SMPE along ``linear” system trajectory is computed as 3.19 p.u. according to the computation strategy as given in Ref. \cite{6}.
This is quite different from the superimposed-machine critical transient energy along “actual” critical stable system trajectory (2.79 p.u.) as in Table \ref{table5}. 
\par From all the analysis above, one question emerges: Is the involvement of the UEP in TSA reasonable?
\par In order to answer this question, the procedures of the UEP-based SMPE are re-described as below.
\vspace*{0.5em}
\\ (i) The UEP is fictionally created because it does not exist along actual post-fault system trajectory \cite{10}.
\\ (ii) The linear system trajectory is fictional because UEP is fictional.
\\ (iii) The SMPE at UEP is computed along the linear system trajectory.
\\
\par (i) to (iii) further indicate the following two deductions
\vspace*{0.5em}
\\ (Deduction-I) Because both the UEP and the linear system trajectory are fictional, the two fictional concepts will further bring the ``computation errors” of the superimposed-machine based critical transient energy, as analyzed in Section \ref{section_VIIB}.
\\ (Deduction-II) Further, the superimposed-machine based critical transient energy is not the real critical transient energy of the original system, because the superimposed machine is pseudo, as analyzed in Section \ref{section_VIIA}.
\vspace*{0.5em}
\par Following the two deductions, it is clear that the use of UEP in the computation of superimposed-machine critical transient energy with mistake will become ``one mistake with a far greater mistake”. This is also the third defect of UEP.

\section{CONCLUSIONS} \label{section_VIII}
In this paper the mechanisms of the superimposed machine and its defects are analyzed. Due to the global monitoring of the original system trajectory, the two-machine system is unstable to be modeled, and the SMTE is mistakenly defined in a superimposed manner. Against this background, the superimposed machine becomes a pseudo machine without equation of motion. The transient characteristics of the superimposed machine are explained from the individual-machine perspective. It is clarified that the pseudo superimposed machine completely violates all the machine paradigms.
The violations bring the two inherit defects of the superimposed-machine method in TSA: (i) the stability of the superimposed machine is unable to be characterized precisely, and (ii) the variance of the original system trajectory is unstable to be depicted clearly. The two defects are fully reflected in the definitions of the superimposed-machine based transient stability concepts. It is found that the swing and the critical stability of the system are defined quite confusing, while the SMPES is also modeled with the pseudo NEC characteristic.
In the end of the paper, the third defect of UEP is exposed. It is clarified that the UEP seems to be an individual-machine based concept, yet it actually serves the computation of the superimposed machine based critical transient energy. This is a ``one mistake with a far greater mistake”.
From the analysis in this paper, all the defects and problems of the superimposed machine in TSA is essentially caused by the global monitoring and the corresponding superimposition.
\par  Aiming to solve the failure of NEC in the SMTE that has residual SMKE problem, modern global analysts attempted to separate all machines in the system into two ``groups”. 
After that, the two equivalent machines are established through the equivalence of all machines in the two groups. In this way the well-known equivalent machine method is formed. The equivalent machine strictly follows the machine paradigms and thus the residual SMKE problem is completely solved.
Against this background, both the stability characterization advantage and the trajectory depiction advantage are preserved in the equivalent-machine based TSA. However, because the original system is replaced with the equivalent system, the equivalent machine will show differences from the individual machine. This will be analyzed in the companion paper.

%

%
%
%




\end{document}